\documentclass[journal,twoside]{IEEEtran}
\usepackage{graphicx}
\usepackage[cmex10]{amsmath}
\usepackage{amssymb,amsfonts,amsthm}
\usepackage{bm}
\usepackage{cite}
\usepackage{url}
\usepackage{algorithm,algorithmicx,algpseudocode}
\usepackage{dblfloatfix}
\usepackage{placeins}
\usepackage{array}
\usepackage{color}

\DeclareMathOperator{\diag}{diag}
\newcolumntype{L}[1]{>{\raggedright\arraybackslash}p{#1}}

\newcommand{\R}{\mathbb{R}}
\newcommand{\C}{\mathbb{C}}
\newcommand{\PFC}{\textrm{P}}
\newcommand{\VC}{\textrm{V}}
\newcommand{\AGC}{\textrm{A}}
\newcommand{\net}{\textrm{net}}
\newcommand{\act}{\textrm{act}}
\newcommand{\refv}{\textrm{ref}}

\allowdisplaybreaks[4]

\begin{document}
\title{Optimal Coordination of Multiple Ancillary \\Service Provisions in Active Distribution Networks}
\author{Xinbao Jia,~\IEEEmembership{Student Member,~IEEE,}
Yi Guo,~\IEEEmembership{Member,~IEEE,}
Ognjen Stanojev,~\IEEEmembership{Member,~IEEE,}\\
Zhen Chen,~\IEEEmembership{Member,~IEEE,}
Yuning Jiang,~\IEEEmembership{Member,~IEEE,}
and Jian Sun,~\IEEEmembership{Senior Member,~IEEE}
\thanks{This work was supported in part by the National Natural Science Foundation of China under grant No. 62495090, 62495095,  62688201, and 62603057. This work was also supported in part by the Fundamental Research Funds for the Central Universities. \textit{(Corresponding author: Yi Guo)}}
\thanks{X. Jia, Y. Guo, Z. Chen, and J. Sun are with the School of Automation, Beijing Institute of Technology, Beijing, China.  (email: \{xinbao.jia, yi.guo, chenzhen76, sunjian\}@bit.edu.cn)}
\thanks{O. Stanojev is with ABB Corporate Research Center, Switzerland. (email: ognjen.stanojev@ch.abb.com)
}
\thanks{Y. Jiang is with the Institute for Transport Planning and Systems at ETH Zürich and also with the Automatic Control Laboratory at EPFL, Switzerland. (email: yuning.jiang@ieee.org)}
}
\maketitle

\setlength\abovedisplayskip{4pt}
\setlength\belowdisplayskip{4pt}

\begin{abstract}
We propose a multi-rate model predictive control (MPC) framework for coordinating primary frequency control (PFC), automatic generation control (AGC), and voltage control (VC) in active distribution networks. The framework allocates distributed energy resource (DER) flexibility among these three ancillary services while retaining their different update rates. To account for the intertemporal flexibility of electric vehicle aggregations and aggregations of buildings with thermostatically controlled loads (TCLs), we incorporate time-varying power–energy envelopes into our multi-rate MPC formulation. These constraints couple the combined PFC and AGC response to the demand-side flexibility available later in the prediction horizon. Furthermore, the input-multirate structure enables coordinated ancillary service provision by DERs with different response dynamics. The controller optimizes the three services together, allowing fast regulation to be coordinated with the developing slow response within the available DER flexibility. We demonstrate the coordinated provision of the three ancillary services through closed-loop simulations on a modified IEEE 33-bus system using a nonlinear differential-algebraic model. 
\end{abstract}

\begin{IEEEkeywords}
active distribution networks, ancillary services, demand-side flexibility, model predictive control, multi-rate control, electric vehicles, thermostatically controlled loads
\end{IEEEkeywords}

\section{Introduction}\label{sec:introduction}
\IEEEPARstart{T}{he} transition toward renewable-dominated power systems presents new challenges for frequency regulation and voltage support. As converter-interfaced renewable generation continues to replace conventional synchronous units, the availability of conventional resources providing these ancillary services decreases~\cite{milano2018}. This motivates ancillary service provision by alternative resources. Meanwhile, the growing deployment of distributed energy resources (DERs), including dispatchable generators, battery energy storage systems (BESSs), electric vehicles (EVs), and thermostatically controlled loads (TCLs), offers controllable flexibility in distribution networks (DNs). Their controllability is transforming traditionally passive DNs into active networks with significant operational flexibility~\cite{olivares2014,karagiannopoulos2020,ravi2023}. Although individual DERs generally have limited capacities, their coordinated aggregations can enable active DNs to provide voltage support and frequency regulation to the upstream grid~\cite{miah2026}. 

Existing studies have investigated PFC and AGC using controllable loads and other flexible resources~\cite{kontis2021,tang2018,zhao2014,shen2025}, and voltage support using distributed active power sources and flexible loads~\cite{rousis2021,valverde2019,karagiannopoulos2021,escobar2022}.
Although these studies highlight the capabilities of DERs in active DNs for frequency and voltage regulation, they mainly focus on single ancillary services or one control layer at a time. In the case that the same DER portfolio participates in multiple services, the feasible allocation depends on their joint use of device and network capability~\cite{namor2019,hanif2022}. For example, hierarchical architectures separate slow scheduling from fast regulation~\cite{scattolini2009,guerrero2011}, with fast control actions determined under the allocation set by the supervisory layer. By contrast, centralized MPC can jointly optimize DER setpoints within a common constrained problem, coordinating their use of the available flexibility~\cite{parisio2014}. In fact, jointly coordinating multiple services within a common optimization also requires reconciling their inherently different control time scales.

PFC and VC require fast adjustments, whereas AGC provides a relatively slower response over a longer interval~\cite{rebours2007}. Despite these different time scales, the corresponding services rely on the same underlying DER and network capabilities, including active and reactive power availability, voltage limits, and branch current capacity~\cite{dozein2021,gutierrez2021}. As a result, service schedules that are feasible when determined independently may become incompatible when implemented simultaneously. A further challenge is how to accommodate these different services within a single optimal control framework. Updating services at the fast rate introduces AGC decisions at each fast control interval, which increases the number of optimization variables. In contrast, adopting a common slow update rate reduces the computational burden but restricts PFC and VC to less frequent updates. This will limit their ability to respond promptly to fast system variations. This trade-off motivates a multi-rate control framework that retains the respective update rates of different services while optimally coordinating DER flexibility for the joint provision of multiple ancillary services. 

Multi-rate model predictive control (MPC) accommodates different measurement or control update rates within a finite horizon formulation~\cite{carini1990}. Subsequent studies have extended multi-rate MPC to applications in which measurements or actuators operate at different rates. In networked control, a distributed MPC formulation incorporates state measurements available at different sampling rates and adjusts controller coordination according to the available measurements~\cite{heidarinejad2011}. More recently, a unified multi-rate MPC formulation has been developed for a humanoid robot to coordinate joint and jet engine actuation with different update rates~\cite{gorbani2025}. These applications illustrate how a multi-rate structure accommodates different measurement and actuation requirements within predictive control. The same principle is relevant to ancillary service provision in power systems, where PFC, AGC, and VC have different time scales and response requirements but share the available DER flexibility. Coordinating these services therefore requires both suitable update rates and an allocation that accounts for their combined use of the participating resources.

We propose a centralized multi-rate MPC (C-MPC) framework for coordinating PFC, AGC, and VC in active DNs. The framework jointly allocates available DER flexibility among the three services, accommodating their different update rates and the intertemporal flexibility of demand-side resources. The resulting finite horizon problem admits an equivalent convex reformulation as a quadratic problem for receding horizon implementation. 

The main contributions are summarized as follows.
\begin{itemize}

\item[1)] \textit{Multi-Rate Coordination:}
We first formulate a continuous-time multi-rate optimal control problem and then parameterize the control inputs and discretize the DER dynamics to obtain a finite-horizon MPC formulation. The resulting centralized multi-rate MPC jointly optimizes the PFC, AGC, and VC setpoints on a common time grid, with each service updated at its prescribed rate. DER response models capture how the physical outputs develop following setpoint adjustments, accounting for different device dynamics. The distribution system model then maps these outputs to bus voltages, branch currents, and service delivery. Together with DER capability limits, network operating constraints, and demand-side power--energy constraints, these models define the admissible service allocation over the prediction horizon. Hence, the three services operate at their different update rates while their use of available DER flexibility is coordinated within a single finite-horizon problem.

\item[2)] \textit{Demand-Side Flexibility:}
Demand-side flexibility is characterized not only by available power capacity but also by the duration over which the response can be maintained \cite{cai2026}. Based on aggregate flexibility models \cite{zhao2017,barot2017,wen2024}, we therefore introduce time-varying power--energy envelopes for EV aggregations and aggregations of buildings with TCLs. These envelopes bound aggregate power and cumulative energy deviation from baseline consumption, capturing how current power use affects the flexibility available later in the prediction horizon.

\item[3)] \textit{Comparative Assessment:}
Compared with approaches focused on individual ancillary services \cite{kontis2021,karagiannopoulos2021}, the framework accounts for the interaction among PFC, AGC, and VC through shared resources. Previously proposed hierarchical architectures separate slow scheduling from fast regulation across different control layers \cite{scattolini2009,guerrero2011}. In contrast, C-MPC combines centralized service allocation with an input-multirate structure, allowing fast regulation and sustained support to be coordinated according to the different service requirements and DER response dynamics. This coordination allows the PFC allocation to account for the developing AGC response and their shared use of available DER flexibility. 
\end{itemize}

The rest of the paper is organized as follows. Section~\ref{sec:mr_framework} develops the general multi-rate optimal control framework. Section~\ref{sec:model} presents the ancillary service requests and DER response models. Section~\ref{sec:coordination} introduces the demand-side flexibility and distribution system models, formulates the centralized multi-rate MPC problem, and describes its receding horizon implementation. Section~\ref{sec:results} includes the simulation setup and numerical results, and Section~\ref{sec:conclusion} concludes the paper.

\textit{Notation}. The sets of real and complex numbers, positive integers, and nonnegative integers are denoted by $\mathbb R$, $\mathbb C$, $\mathbb N$, and $\mathbb Z_{\geq 0}$, respectively. We write $\|\mathbf x\|_{\mathbf Q}^2:=\mathbf x^\top\mathbf Q\mathbf x$. The operators $\Re(\cdot)$, $\Im(\cdot)$, and $(\cdot)^*$ denote the real part, imaginary part, and complex conjugation; $\mathrm j$ is the imaginary unit. The operator $\diag(\mathbf x)$ forms a diagonal matrix from $\mathbf x$. The notation $(\cdot)_{k|t}$ denotes a prediction at step $k$ formed at update time $t$, and a hat denotes a measured or estimated quantity. Bold lowercase and uppercase Latin symbols denote vectors and matrices; bold Greek symbols denote vectors or matrices. Scalars are unbolded, and descriptive subscripts and superscripts are upright.

\section{Multi-Rate Optimal Control Framework}
\label{sec:mr_framework}
In this section, we first formulate a continuous-time optimal control problem for ancillary service coordination in active DNs. We then parameterize the control inputs and discretize the prediction model to obtain a finite-horizon multi-rate MPC formulation.

\subsection{Continuous-Time Optimal Control for Active Distribution Networks}

An active DN provides ancillary services by adjusting the active and reactive power setpoints of its DERs. The DERs have continuous-time response dynamics, even though their power setpoints are updated at discrete instants. Their physical outputs may therefore continue to evolve between successive setpoint updates, subject to DER capability and network operating constraints. To predict this evolution, let $\mathbf x(\tau)\in\R^{n_x}$ denote the state of a continuous-time model for the relevant DER dynamics at time $\tau$ and let $\mathbf u(\tau)\in\R^{n_u}$ collect the control setpoints, where $n_x,n_u\in\mathbb N$ denote their dimensions.

The control inputs also need to accommodate the different time scales of the services. PFC and VC require fast setpoint adjustments, whereas AGC provides sustained support with a slower update period. Despite these different update periods, the services share DER power and energy flexibility and the available network capacity. For example, PFC and AGC share the available active power capability, and their combined response affects the demand-side flexibility later in the prediction horizon. The VC allocation is also coupled to these active power services through DER capability limits and network operating constraints. To accommodate the different service update periods, we divide the control input into $M\in\mathbb N$ groups, each with its own update schedule. We write $\mathbf u(\tau)=(\mathbf u^{(1)}(\tau)^\top,\ldots,\mathbf u^{(M)}(\tau)^\top)^\top$ with $\mathbf u^{(i)}(\tau)\in\R^{n_i}$ and $\sum_{i=1}^{M}n_i=n_u$. Each input group consists of the DER setpoints associated with one ancillary service. Over a prediction horizon $T_p\in\R_{>0}$ starting at time $t\in\R_{\geq0}$, we formulate the continuous-time multi-rate optimal control problem:
\begin{subequations}\label{eq:general_ct_mpc}
\begin{align}
\mathcal P_t^{\mathrm{ct}}:\quad \;\underset{\mathbf u(\cdot)}{\min}\quad
&\int_t^{t+T_p}\ell(\mathbf x(\tau),\mathbf u(\tau))\,d\tau,\label{eq:general_ct_objective}\\
\text{subject to}\quad
&\dot{\mathbf x}(\tau)=\mathbf f(\mathbf x(\tau),\mathbf u(\tau)),\label{eq:general_ct_dynamics}\\
&\mathbf x(t)=\hat{\mathbf x}(t),\label{eq:general_ct_initial}\\
&\mathbf g(\mathbf x(\tau),\mathbf u(\tau),\tau)\leq\mathbf0,\label{eq:general_ct_ineq}\\
&\mathbf h(\mathbf x(\tau),\mathbf u(\tau),\tau)=\mathbf0,\label{eq:general_ct_eq}\\
&\mathbf u^{(i)}(\cdot)\in\mathcal U_i^{\mathrm{MR}},\quad i=1,\ldots,M,\label{eq:general_ct_input}
\end{align}
\end{subequations}
where $\mathbf f:\R^{n_x}\times\R^{n_u}\to\R^{n_x}$ describes the continuous-time DER response to active and reactive power setpoints, $\hat{\mathbf x}(t)\in\R^{n_x}$ is the initial DER state obtained from measurements or estimation, and $\ell:\R^{n_x}\times\R^{n_u}\to\R$ is the running cost associated with ancillary service provision. The function $\mathbf g:\R^{n_x}\times\R^{n_u}\times\R_{\geq0}\to\R^{n_g}$ represents DER capability limits, demand-side power--energy bounds, and limits on bus voltages and branch currents. The function $\mathbf h:\R^{n_x}\times\R^{n_u}\times\R_{\geq0}\to\R^{n_h}$ describes the equality relations associated with the distribution system model and ancillary service delivery requirements, with $n_g,n_h\in\mathbb N$. These constraints apply to the combined physical response of the services throughout $\tau\in[t,t+T_p]$. The set $\mathcal U_i^{\mathrm{MR}}$ contains the admissible DER setpoint functions for group $i$ under its prescribed update schedule. The setpoint applied immediately before $t$ is retained until the first allowable update, and each setpoint is held constant between updates.

\subsection{Multi-Rate Input Parameterization and Discretization}

The continuous-time problem \eqref{eq:general_ct_mpc} describes the DER response under service setpoints with different update periods. For digital implementation, we parameterize these setpoints and discretize the DER dynamics to obtain a finite-dimensional optimization problem. Let \(T_s\in\mathbb R_{>0}\) denote the sampling period and \(T_{\mathrm{sh}}\in\mathbb R_{>0}\) denote the shooting interval, with \(T_{\mathrm{sh}}=T_s\).

The MPC problem is solved every \(T_{\mathrm{MPC}}=rT_s\), where \(r\in\mathbb N\), while the setpoints of input group \(i\) are updated every \(T_i=q_iT_s\), where \(q_i\in\mathbb N\). We assume $q_i/r\in\mathbb N$ so that the input updates coincide with MPC update instants. Each input group is parameterized using a zero-order hold with its prescribed update period $T_i$:
\begin{equation}\label{eq:general_hold}
\mathbf u^{(i)}(\tau)=\mathbf u^{(i)}(jT_i),\quad
\tau\in[jT_i,(j+1)T_i),\quad j\in\mathbb Z_{\geq0},
\end{equation}
where $j$ indexes the update intervals of input group $i$. Although the service setpoints are updated at different rates, the DER responses are predicted using a common sampling period $T_s$.

\begin{figure}[!htbp]
\centering
\includegraphics[width=1.05\columnwidth]{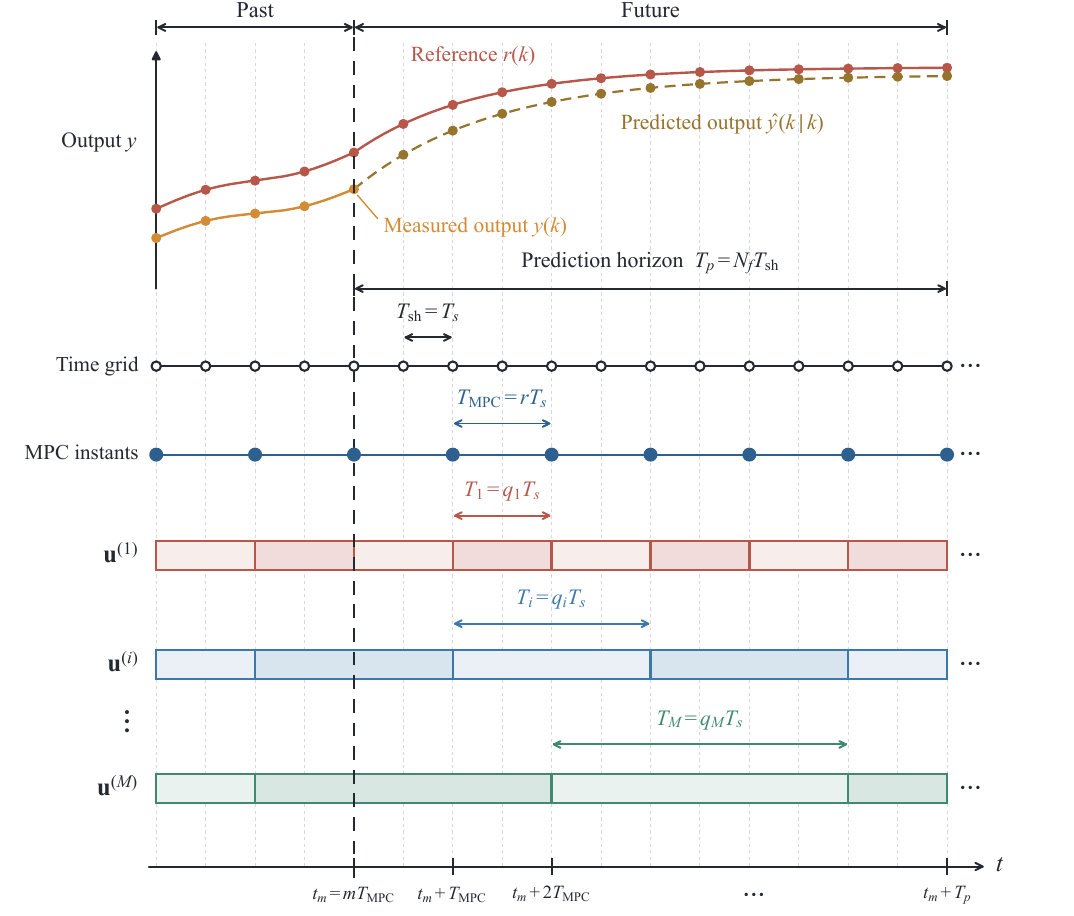}
\caption{Multi-rate MPC timing diagram showing different input update periods on a common time grid.}
\label{fig:timescale}
\end{figure}

At MPC update $t_m=mT_{\mathrm{MPC}}=mrT_s$, $m\in\mathbb Z_{\geq0}$, we consider a prediction horizon $T_p=N_fT_{\mathrm{sh}}$ with $N_f\in\mathbb N$ and index the prediction steps by $\mathcal K:=\{0,\ldots,N_f-1\}$. The state $\mathbf x_{k|t_m}$ corresponds to time $t_m+kT_s$, and $\mathbf u_{k|t_m}^{(i)}\in\R^{n_i}$ is the setpoint applied over $[t_m+kT_s,t_m+(k+1)T_s)$. To express setpoint changes relative to the previously applied values, we define
\begin{subequations}\label{eq:general_increment}
\begin{align}
\Delta\mathbf u_{0|t_m}^{(i)}
&=\mathbf u_{0|t_m}^{(i)}-\mathbf u^{(i)}(t_m^-),\label{eq:general_input_initial}\\
\Delta\mathbf u_{k+1|t_m}^{(i)}
&=\mathbf u_{k+1|t_m}^{(i)}
-\mathbf u_{k|t_m}^{(i)},\quad k=0,\ldots,N_f-2.\label{eq:general_input_delta}
\end{align}
\end{subequations}
The prescribed update schedule is enforced by
\begin{equation}\label{eq:general_update_constraint}
\Delta\mathbf u_{k|t_m}^{(i)}=\mathbf0,\quad
(mr+k)\bmod q_i\ne0,\quad k\in\mathcal K.
\end{equation}
Here, $mr+k$ is the sampling index corresponding to time $t_m+kT_s$. This constraint ensures that the predicted setpoints can change only at the prescribed update instants as the prediction horizon shifts. Fig.~\ref{fig:timescale} illustrates how the prediction sampling period, shooting interval, MPC execution period, and input update periods are defined and related.

Under the zero-order-hold input parameterization, we discretize the continuous-time DER dynamics in \eqref{eq:general_ct_dynamics} with sampling period $T_s$. For each prediction step, the continuous-time model is solved over $T_s$ with initial state $\mathbf x_{k|t_m}$ and constant input $\mathbf u_{k|t_m}$. The resulting discrete-time prediction model is
\begin{subequations}\label{eq:general_discrete_dynamics}
\begin{align}
\mathbf x_{k+1|t_m}
&=\mathbf F_{T_s}(\mathbf x_{k|t_m},\mathbf u_{k|t_m}),
\label{eq:general_discrete_state}\\
\mathbf x_{0|t_m}
&=\hat{\mathbf x}(t_m).
\label{eq:general_discrete_initial}
\end{align}
\end{subequations}
where $\mathbf F_{T_s}:\R^{n_x}\times\R^{n_u}\to\R^{n_x}$ denotes the state transition map over one sampling period.

The integral cost is approximated by a Riemann sum, while the operating constraints are discretized over the prediction horizon. Let $\mathbf g_{k|t_m}$ and $\mathbf h_{k|t_m}$ denote the corresponding constraint functions at prediction step $k$. Together with the discrete-time prediction model and the multi-rate input structure, these approximations yield the following finite-horizon MPC problem:
\begin{subequations}\label{eq:general_discrete_mpc}
\begin{align}
\mathcal P_{t_m}^{\mathrm{MR}}:\quad \;
\underset{\mathbf U_{t_m}}{\min}\quad
&T_s\sum_{k\in\mathcal K}
\ell(\mathbf x_{k|t_m},\mathbf u_{k|t_m}),
\label{eq:general_mr_objective}\\
\text{subject to}\quad
&\mathbf x_{0|t_m}=\hat{\mathbf x}(t_m),
\label{eq:general_mr_initial_state}\\
&\mathbf u_{0|t_m}^{(i)}=\mathbf u^{(i)}(t_m^-)+\Delta\mathbf u_{0|t_m}^{(i)},
\label{eq:general_mr_initial_input}\\
&\mathbf x_{k+1|t_m}
=\mathbf F_{T_s}(\mathbf x_{k|t_m},\mathbf u_{k|t_m}),
\label{eq:general_mr_state}\\
&\Delta\mathbf u_{k+1|t_m}^{(i)}
=\mathbf u_{k+1|t_m}^{(i)}-\mathbf u_{k|t_m}^{(i)},
\label{eq:general_mr_increment}\\
&\mathbf g_{k|t_m}(\mathbf x_{k+1|t_m},\mathbf u_{k|t_m})
\leq\mathbf0,
\label{eq:general_mr_ineq}\\
&\mathbf h_{k|t_m}(\mathbf x_{k+1|t_m},\mathbf u_{k|t_m})
=\mathbf0,
\label{eq:general_mr_eq}\\
&\Delta\mathbf u_{k|t_m}^{(i)}=\mathbf0,\;
(mr{+}k)\bmod q_i\ne0,
\label{eq:general_mr_hold}
\end{align}
\end{subequations}
where $\mathbf U_{t_m}:=(\mathbf u_{0|t_m}^{\top},\ldots,\mathbf u_{N_f-1|t_m}^{\top})^{\top}\in\R^{N_fn_u}$ collects the control inputs over the prediction horizon. We solve problem~\eqref{eq:general_discrete_mpc} at each MPC update and apply the first optimized setpoints of the input groups scheduled to update. The remaining input groups keep their previously applied setpoints. At the next MPC update, the prediction horizon is shifted and the problem is solved again using the updated operating point and states.

This formulation coordinates service setpoints with different update periods while predicting their combined physical response on a common time grid. In the next section, we apply this framework to PFC, AGC, and VC by specifying the corresponding input groups, control update periods, service requests, and DER response models.

\section{Ancillary Service Requests and DER Response Models}\label{sec:model}
\begin{figure}[!htbp]
\centering
\includegraphics[width=\columnwidth]{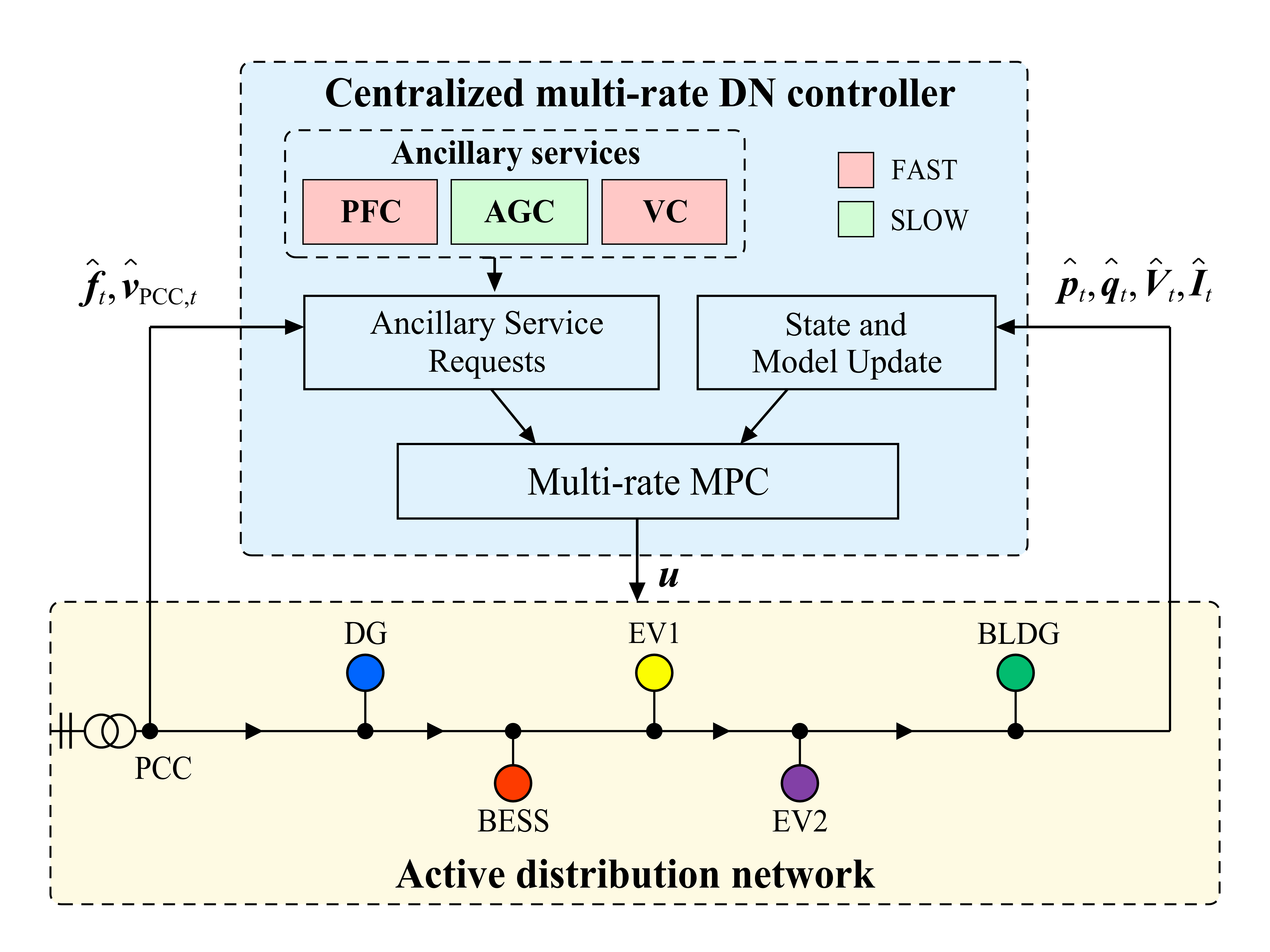}
\caption{Proposed multi-rate control structure for an active DN with demand-side flexibility. PFC and VC operate at the fast update rate, whereas AGC operates at the slow update rate. The MPC uses ancillary service requests and updated states to compute DER setpoints. Feeder and DER measurements update the operating point and aggregate energy states. The arrows along the feeder indicate the branch reference directions from the upstream bus toward downstream buses.}
\label{fig:framework}
\end{figure}

In this section, we develop the models required for the multi-rate MPC formulation to coordinate multiple ancillary services. Specifically, we introduce the multi-rate optimal control structure with ancillary service requests, followed by the DER response models. Fig.~\ref{fig:framework} provides an overview of the proposed framework and its main information flows.

\subsection{Ancillary Service Requests}\label{sec:service_requests}
We consider a heterogeneous portfolio of DERs including diesel generators (DGs), BESSs, EV charging demand aggregations, and aggregations of buildings with TCLs. Each EV or TCL aggregation represents a group of devices controlled collectively for ancillary service provision. These resources provide frequency support by adjusting their active power outputs or consumption. Let $\mathcal D_{\textrm{DG}}$, $\mathcal D_{\textrm{BESS}}$, $\mathcal D_{\textrm{EV}}$, and $\mathcal D_{\textrm{BLDG}}$ denote the corresponding resource sets, respectively, with $\mathcal D$ denoting their union. Let $n_d:=|\mathcal D|\in\mathbb N$ denote the number of DERs. We use $\mathcal D_Q\subseteq\mathcal D$ to collect the DERs as a subset to include devices that have the capability to provide reactive power support.

Because ancillary services operate at different time scales, we specify the multi-rate time structure used in the subsequent control formulation. PFC and VC use the fast setpoint update period $T_f$, and AGC uses the period $T_a=qT_f$, where $q\in\mathbb N$. To apply the discrete multi-rate framework developed in Section~\ref{sec:mr_framework}, we choose the sampling period and MPC execution period as $T_s=T_{\mathrm{MPC}}=T_f$, corresponding to $r=1$ in the general framework.

We coordinate the three services over a finite prediction horizon starting at the current update time $t_m$. PFC and AGC regulate active power, whereas VC regulates reactive power. Each service corresponds to one input group, giving $M=3$.

For $\nu\in\{\PFC,\VC,\AGC\}$, let $\mathbf u^\nu_{k|t_m}\in\R^{n_d}$ denote the setpoint vector for the DERs providing the corresponding service. These vectors follow the DER ordering defined by $\mathcal D$, with zero entries for resources not participating in the corresponding service. The control input in \eqref{eq:general_discrete_mpc} is then $\mathbf u_{k|t_m}=((\mathbf u^{\PFC}_{k|t_m})^\top,(\mathbf u^{\VC}_{k|t_m})^\top,(\mathbf u^{\AGC}_{k|t_m})^\top)^\top\in\R^{3n_d}$. The input update ratios are $q_{\PFC}=q_{\VC}=1$ and $q_{\AGC}=q$. The corresponding setpoint increments $\Delta\mathbf u^\nu_{k|t_m}\in\R^{n_d}$ are defined by \eqref{eq:general_increment}.

The PFC and VC setpoints can change at every prediction step. For AGC, \eqref{eq:general_update_constraint} permits a change only at prediction steps satisfying $(m+k)\bmod q=0$. We collect these steps in $\mathcal A_m:=\{k\in\mathcal K:(m+k)\bmod q=0\}$ and impose $\Delta\mathbf u^{\AGC}_{k|t_m}=\mathbf0$ for $k\in\mathcal K\setminus\mathcal A_m$. Choosing $N_f$ as a multiple of $q$ gives $N_a:=N_f/q$ allowable AGC updates within the prediction horizon, with $|\mathcal A_m|=N_a$. Thus, all three services are represented on the common time grid, with their setpoints updated at the prescribed rates. 

At each fast control update, the PFC and VC requests are determined from the deviations of the measured system frequency and PCC voltage from the respective reference. We define
\begin{subequations}\label{eq:service_errors}
\begin{align}
e_{t_m}^f&:=f_0-\hat f_{t_m},\label{eq:frequency_error}\\
e_{t_m}^{\VC}&:=v_{\mathrm{PCC}}^{\refv}-\hat v_{\mathrm{PCC},t_m}.\label{eq:voltage_error}
\end{align}
\end{subequations}
Here, $\hat f_{t_m}$ and $\hat v_{\mathrm{PCC},t_m}$ denote the measured system frequency and PCC voltage magnitude, respectively, while $f_0$ and $v_{\mathrm{PCC}}^{\refv}$ are their reference values. PCC is located at the receiving end of the main feeder. With this sign convention, positive $e_{t_m}^{f}$ and $e_{t_m}^{\VC}$ correspond to frequency and voltage values below their references, respectively. These error signals are then used to construct the aggregate PFC and VC requests.

Using the frequency control error in \eqref{eq:frequency_error}, we consider a bounded aggregated PFC request for the DN. Given the prescribed lower and upper PFC limits $\underline P^{\PFC},\overline P^{\PFC}\in\R$, respectively, the proportional gain $K_P\in\R_{>0}$ maps the frequency error to the aggregate PFC request:
\begin{equation}\label{eq:pfc_request}
\rho_{t_m}^{\PFC}=\begin{cases}
\min\{\overline P^{\PFC},K_Pe_{t_m}^f\},&e_{t_m}^f\geq0,\\
\max\{\underline P^{\PFC},K_Pe_{t_m}^f\},&e_{t_m}^f<0,
\end{cases}
\end{equation}
Here, $\rho^{\PFC}_{t_m}$ denotes the aggregate PFC support requested at the PCC. While PFC provides the fast proportional response to the frequency deviation, AGC provides a slower integral correction for frequency restoration.

For the single-area equivalent considered here, the AGC request is determined from the frequency deviation \eqref{eq:frequency_error} without a tie line power term. We model the continuous request from the upstream AGC controller using integral control \cite{kerci2021}:
\begin{equation}\label{eq:agc_request_model}
\rho^{\AGC}(\tau)=\rho^{\AGC}(0)
+K_{i,A}\int_0^\tau e^f(s)\,\mathrm ds,
\end{equation}
where $K_{i,A}$ is the integral gain and $\rho^{\AGC}(0)$ is the given initial request. The continuous frequency error is $e^f(\tau):=f_0-f(\tau)$, with $f(\tau)$ denoting the system frequency.

The AGC request may change between successive DER setpoint updates. To provide current demand information at each MPC solve, we sample this continuous request at the fast period $T_f$:
\begin{equation}\label{eq:agc_request_sampling}
\rho_{t_m}^{\AGC}:=\rho^{\AGC}(t_m),\qquad t_m=mT_f.
\end{equation}
Each sampled value is held until the next sample. The DER AGC setpoints are updated every $T_a$ according to \eqref{eq:general_update_constraint}.

Similarly, the VC request is built from the PCC voltage error in \eqref{eq:voltage_error}. Given the prescribed lower and upper VC request limits $\underline Q^{\VC},\overline Q^{\VC}\in\R$, respectively, the gain $K_V\in\R_{>0}$ maps the PCC voltage error to the aggregate VC request:
\begin{equation}\label{eq:vc_request}
\rho_{t_m}^{\VC}=\begin{cases}
\min\{\overline Q^{\VC},K_Ve_{t_m}^{\VC}\},&e_{t_m}^{\VC}\geq0,\\
\max\{\underline Q^{\VC},K_Ve_{t_m}^{\VC}\},&e_{t_m}^{\VC}<0.
\end{cases}
\end{equation}
Here, $\rho_{t_m}^{\VC}$ represents the desired aggregate reactive power support at the PCC. At each optimization, the current PFC, VC, and sampled AGC requests are held fixed over the prediction horizon and refreshed at the next MPC update.

\subsection{DER Response Models and Physical Outputs}
The setpoint trajectories defined above specify the DER response for different ancillary services. However, due to their inherent dynamics, the physical outputs of different DERs may not track these setpoints instantaneously. We therefore employ low-order discrete-time models to capture the dominant response dynamics while keeping a compact subsequent control formulation.

To specify the prediction model in \eqref{eq:general_discrete_mpc}, let $n_\nu\in\mathbb N$ denote the number of response states for $\nu$, and let $\mathbf x_{k|t_m}^{\nu}\in\R^{n_\nu}$ and $\mathbf r_{k|t_m}^{\nu}\in\R^{n_d}$ denote the internal response state and the corresponding physical output of service $\nu$, respectively. The output vectors follow the DER ordering defined by $\mathcal D$, with zero entries for resources not participating in the corresponding service. The response is represented by the discrete-time state-space model
\begin{subequations}\label{eq:device_model}
\begin{align}
\mathbf x_{k+1|t_m}^{\nu}
&=\mathbf A_{t_m}^{\nu}\mathbf x_{k|t_m}^{\nu}
+\mathbf B_{t_m}^{\nu}\mathbf u_{k|t_m}^{\nu},
\label{eq:device_state}\\
\mathbf r_{k|t_m}^{\nu}
&=\mathbf C_{t_m}^{\nu}\mathbf x_{k|t_m}^{\nu}
+\mathbf D_{t_m}^{\nu}\mathbf u_{k|t_m}^{\nu},
\quad k\in\mathcal K.
\label{eq:device_output}
\end{align}
\end{subequations}
Here, $\mathbf A_{t_m}^{\nu}\in\R^{n_\nu\times n_\nu}$, $\mathbf B_{t_m}^{\nu}\in\R^{n_\nu\times n_d}$, $\mathbf C_{t_m}^{\nu}\in\R^{n_d\times n_\nu}$, and $\mathbf D_{t_m}^{\nu}\in\R^{n_d\times n_d}$ denote the matrices in the response model, discretized with the common period $T_s$. The output $\mathbf r_{k|t_m}^{\nu}$ denotes the physical response at the end of prediction step $k$, expressed in terms of the state and setpoint at the beginning of that step. At update time $t_m$, we initialize the response state as $\mathbf x_{0|t_m}^{\nu}=\hat{\mathbf x}_{t_m}^{\nu}$, where $\hat{\mathbf x}_{t_m}^{\nu}$ is determined from the current DER outputs and available device states. When the required device states are not directly measured, they may be estimated from the available measurements \cite{rao2003,liu2021}. The physical output immediately before the update is denoted by $\hat{\mathbf r}_{t_m}^{\nu}$.

In the device model formulation below, we omit the prediction origin suffix $|t_m$ for readability. Let $d$ denote a generic resource index. For each diesel generator $g\in\mathcal D_{\textrm{DG}}$, we use a first-order model to represent its response to the service adjustment:
\begin{subequations}\label{eq:dg_dynamics}
\begin{align}
x^{\nu}_{g,j+1}
&=\alpha_g^{\nu}x^{\nu}_{g,j}
+(1-\alpha_g^{\nu})u^{\nu}_{g,j},
\label{eq:dg_active_dynamics}\\
r^{\nu}_{g,j}&=x^{\nu}_{g,j+1},
\quad j\in\mathcal K.
\label{eq:dg_reactive_dynamics}
\end{align}
\end{subequations}
where $T_g^{\nu}>0$ is the response time constant of generator $g$ for service $\nu$, and $\alpha_g^{\nu}=\exp(-T_s/T_g^{\nu})$.

For BESSs and EV aggregations, the converter dynamics are assumed to be sufficiently fast relative to the corresponding service update intervals. Their physical outputs are therefore assumed to track the setpoints:
\begin{equation}
r^{\nu}_{d,j}=u^{\nu}_{d,j},\qquad
d\in\mathcal D_{\textrm{BESS}}\cup\mathcal D_{\textrm{EV}},
\quad j\in\mathcal K,
\label{eq:direct_converter_output}
\end{equation}
For each TCL of building aggregations $h\in\mathcal D_{\textrm{BLDG}}$, we represent its PFC and AGC responses also by a first-order dynamic model:
\begin{subequations}\label{eq:build_dynamics}
\begin{align}
x_{h,j+1}^{\nu}
&=\alpha_h^{\nu}x_{h,j}^{\nu}
+(1-\alpha_h^{\nu})u_{h,j}^{\nu},
\label{eq:build_state}\\
r_{h,j}^{\nu}
&=x_{h,j+1}^{\nu},
\quad j\in\mathcal K.
\label{eq:build_output}
\end{align}
\end{subequations}
where $\nu\in\{\PFC,\AGC\}$, $\alpha_h^{\nu}=\exp(-T_s/T_h)$, and $T_h>0$ is the equivalent response time constant.

For notational simplicity, we use $t:=t_m$ to denote the current MPC update time in the following formulation. Because the measured operating point already includes the physical service responses before the update, we define the predicted changes in active and reactive power relative to the corresponding pre-update outputs:
\begin{subequations}\label{eq:physical_channel_changes}
\begin{align}
\Delta\mathbf p_{k|t}^{\PFC,\act}
&=\mathbf r_{k|t}^{\PFC}-\hat{\mathbf r}_t^{\PFC},
&& k\in\mathcal K, \label{eq:pfc_active_output}\\
\Delta\mathbf q_{k|t}^{\VC,\act}
&=\mathbf r_{k|t}^{\VC}-\hat{\mathbf r}_t^{\VC},
&& k\in\mathcal K, \label{eq:vc_reactive_output}\\
\Delta\mathbf p_{k|t}^{\AGC,\act}
&=\mathbf r_{k|t}^{\AGC}-\hat{\mathbf r}_t^{\AGC},
&& k\in\mathcal K. \label{eq:agc_active_output}
\end{align}
\end{subequations}

Let $\hat{\mathbf p}_t\in\R^{n_d}$ and $\hat{\mathbf q}_t\in\R^{n_d}$ denote the measured active and reactive power operating point vectors. We then obtain the total predicted device outputs as
\begin{subequations}\label{eq:physical_output_composition}
\begin{align}
\mathbf p_{k|t}
&=\hat{\mathbf p}_t+\Delta\mathbf p_{k|t}^{\PFC,\act}
+\Delta\mathbf p_{k|t}^{\AGC,\act}, \label{eq:total_active_output}\\
\mathbf q_{k|t}
&=\hat{\mathbf q}_t+\Delta\mathbf q_{k|t}^{\VC,\act},
\qquad k\in\mathcal K. \label{eq:total_reactive_output}
\end{align}
\end{subequations}
These predicted DER outputs are subsequently used to impose the device capability and network constraints.

The BESS supports charging and discharging as well as reactive power regulation. We represent this instantaneous power capability by separate active and reactive power bounds. For each BESS $b\in\mathcal D_{\textrm{BESS}}$, the predicted outputs satisfy
\begin{subequations}\label{eq:bess_capability}
\begin{align}
\underline p_b
&\leq p_{b,k|t}\leq\overline p_b,\\
\underline q_b
&\leq q_{b,k|t}\leq\overline q_b,
\qquad k\in\mathcal K,
\end{align}
\end{subequations}
where $\underline p_b$ and $\overline p_b$ are the lower and upper active power limits, and $\underline q_b$ and $\overline q_b$ are the corresponding reactive power limits.

The DG operating region is restricted by its apparent power limit and minimum active power output. To express these operating limits as linear inequalities, we use a polygonal approximation for each DG $g\in\mathcal D_{\textrm{DG}}$:
\begin{equation}\label{eq:dg_capability}
\mathbf A_g^{\mathrm{cap}}
\begin{bmatrix}
p_{g,k|t}\\
q_{g,k|t}
\end{bmatrix}
\leq\mathbf b_{g,k|t}^{\mathrm{cap}},
\qquad k\in\mathcal K,
\end{equation}
where $\mathbf A_g^{\mathrm{cap}}\in\R^{m_g\times2}$ and $\mathbf b_{g,k|t}^{\mathrm{cap}}\in\R^{m_g}$ define the DG capability polygon with $m_g\in\mathbb N$ facets.

We collect the instantaneous capability limits of all DERs in the polyhedral set
\begin{equation}
\label{eq:instantaneous_capability_set}
\hspace{-3mm}\mathcal X^P_{k|t}:=\biggl\{
(\mathbf p_{k|t},\mathbf q_{k|t}):\;
\mathbf A_d^{\mathrm{cap}}
\begin{bmatrix}
p_{d,k|t}\\q_{d,k|t}
\end{bmatrix}
\leq\mathbf b_{d,k|t}^{\mathrm{cap}}, d\in\mathcal D
\biggr\},
\end{equation}
where $\mathbf A_d^{\mathrm{cap}}\in\R^{m_d\times2}$ is the capability constraint matrix and $\mathbf b_{d,k|t}^{\mathrm{cap}}\in\R^{m_d}$ is the corresponding limit vector for resource $d$ at prediction step $k$, with $m_d\in\mathbb N$ facets.

For the EV and TCL aggregations, these instantaneous limits
are supplemented by the time-varying power-energy envelopes introduced
in the next section.

\section{Centralized Multi-Rate MPC for Ancillary Services}\label{sec:coordination}
In this section, we formulate the centralized multi-rate MPC problem by combining the DER response models with demand-side flexibility and distribution system constraints. The power--energy flexibility envelopes and distribution system model are introduced first, followed by the joint optimization and its receding horizon implementation.

\subsection{Demand-Side Power--Energy Flexibility}
To capture the intertemporal flexibility of EV charging/discharging and building thermal load aggregations without explicitly modeling all individual mobility and thermal states, we represent their aggregated flexibility using time-varying power-energy envelopes relative to their baseline consumption. Since we define EV discharging as positive, whereas building thermal power load is positive for consumption, let $\chi_r=-1$ for an EV aggregation and $\chi_r=1$ for a TCL aggregation, where $r\in\mathcal D_{\textrm{EV}}\cup\mathcal D_{\textrm{BLDG}}$. These factors convert both device conventions to positive consumption.

The variables $P^L_{r,k|t}:=\chi_r p_{r,k|t}$ and $P^{L,0}_{r,k|t}:=\chi_r p^0_{r,k|t}$ denote the controlled and baseline electrical consumption, respectively. Their difference $\Delta P^L_{r,k|t}:=P^L_{r,k|t}-P^{L,0}_{r,k|t}$ represents the deviation from the baseline, where $p^0_{r,k|t}$ denotes the power trajectories without flexibility activation requests. The variable $E_{r,k|t}$ denotes the cumulative electrical-energy deviation from the baseline.

Let $\mathcal I_r$ denote the set of individual devices belonging to aggregation $r$, with $n_r=|\mathcal I_r|\in\mathbb N$. For each $i\in\mathcal I_r$, $P^L_{i,k|t}$ and $P^{L,0}_{i,k|t}$ denote controlled and baseline electrical consumption, respectively, and $E_{i,k|t}$ denotes the cumulative energy deviation from the baseline. The corresponding aggregate power and energy quantities are
\begin{subequations}\label{eq:demand_aggregation}
\begin{align}
P^L_{r,k|t}&=\sum_{i\in\mathcal I_r}P^L_{i,k|t}, \label{eq:aggregate_power}\\
P^{L,0}_{r,k|t}&=\sum_{i\in\mathcal I_r}P^{L,0}_{i,k|t}, \notag\\
E_{r,k|t}&=\sum_{i\in\mathcal I_r}E_{i,k|t}, \label{eq:aggregate_energy}
\end{align}
\end{subequations}
where the individual energy deviation evolves as
\begin{equation}
E_{i,k+1|t}=E_{i,0|t}+\sum_{j=0}^{k}T_s
\bigl(P^L_{i,j|t}-P^{L,0}_{i,j|t}\bigr). \label{eq:individual_energy_recursion}
\end{equation}
Let $\hat E_{r,t}$ denote the cumulative energy deviation of aggregation $r$ at the current update. Over the prediction horizon, the aggregate power and energy deviations satisfy
\begin{subequations}\label{eq:demand_energy}
\begin{align}
E_{r,0|t} & = \hat E_{r,t}, \label{eq:energy_initial}\\
E_{r,k+1|t} & = E_{r,k|t}+T_s\,\Delta P^L_{r,k|t}, \label{eq:energy_recursion}
\end{align}
\end{subequations}
together with the time-varying power and energy bounds
\begin{subequations}\label{eq:demand_energy_model}
\begin{align}
\underline P^L_{r,k|t} & \leq P^L_{r,k|t}\leq\overline P^L_{r,k|t}, \label{eq:power_envelope}\\
\underline E_{r,k+1|t} & \leq E_{r,k+1|t}\leq\overline E_{r,k+1|t}. \label{eq:energy_envelope}
\end{align}
\end{subequations}
Here, $\underline P^L_{r,k|t}$ and $\overline P^L_{r,k|t}$ are the lower and upper consumption limits, while $\underline E_{r,k+1|t}$ and $\overline E_{r,k+1|t}$ denote the corresponding bounds on cumulative energy deviation. These constraints are imposed for $k\in\mathcal K$ and each $r\in\mathcal D_{\textrm{EV}}\cup\mathcal D_{\textrm{BLDG}}$.

For an EV aggregation, we construct the power-energy envelope from vehicle connection status, charging and discharging limits, state of charge (SOC), and departure energy requirements \cite{zhang2017,pertl2019}. Let $a_{i,k|t}\in\{0,1\}$ indicate whether the $i$-th EV is connected at prediction step $k$, and $\epsilon_{i,k|t}\in[0,1]$ denote its SOC. The feasible operation of each EV is described by
\begin{subequations}\label{eq:individual_ev_envelope}
\begin{align}
-a_{i,k|t}\overline P_i^{d}&\leq P^L_{i,k|t}\leq a_{i,k|t}\overline P_i^{c}, \label{eq:individual_ev_power}\\
\epsilon_{i,k+1|t}&=\epsilon_{i,k|t}+{T_s}\frac{P^L_{i,k|t}}{E_i^{\mathrm{cap}}}, \label{eq:individual_ev_energy}\\
\underline\epsilon_i\leq\epsilon_{i,k+1|t}&\leq\overline\epsilon_i,\qquad
\epsilon_{i,k_i^d|t}\geq\epsilon_i^{\mathrm{req}}, \label{eq:individual_ev_soc}
\end{align}
\end{subequations}
Here, $E_i^{\mathrm{cap}}>0$ denotes the battery energy capacity of EV $i$. The initial SOC $\epsilon_{i,0|t}$ is given. The parameters $\overline P_i^c$ and $\overline P_i^d$ denote the nonnegative charging and discharging power limits, respectively, while $\underline\epsilon_i$ and $\overline\epsilon_i$ are the lower and upper bounds of SOC. The parameter $\epsilon_i^{\mathrm{req}}\in[0,1]$ specifies the minimum SOC required at the departure step $k_i^d\in\mathbb N$. Unidirectional charging is obtained by setting $\overline P_i^d=0$. The energy deviation state can equivalently be written as $E_{i,k|t}=E_i^{\mathrm{cap}}(\epsilon_{i,k|t}-\epsilon^0_{i,k|t})$, where $\epsilon^0_{i,k|t}$ denotes the SOC at step $k$ under the baseline consumption profile. Accordingly, we sum up the individual energy deviation bounds across the EVs in an aggregation to give the corresponding aggregate energy-deviation bounds, which, together with the aggregate power bounds, form the EV power--energy flexibility envelope.

The flexibility of a building's thermostatically controlled load aggregation arises from building thermal inertia and the allowable variation of indoor temperature within comfort limits. For building $i$, let $\vartheta_{i,k|t}$ denote the indoor temperature, $\vartheta^a_{k|t}$ the ambient temperature, $H_i$ the heat-loss coefficient, and $C_i$ the thermal capacitance. We use a first-order resistance--capacitance (RC) thermal model to characterize the building response and construct the corresponding flexibility envelope:
\begin{subequations}
\label{eq:individual_build_envelope}
\begin{align}
\vartheta_{i,k+1|t}&=\alpha_i\vartheta_{i,k|t} \notag\\
&\quad +(1-\alpha_i)
\left(\vartheta^a_{k|t}+\frac{\mathrm{COP}_{i,k}P^L_{i,k|t}}{H_i}\right), \label{eq:individual_build_temperature}\\
0\leq P^L_{i,k|t}&\leq\overline P_i,\qquad
\underline\vartheta_i\leq\vartheta_{i,k+1|t}\leq\overline\vartheta_i.\label{eq:individual_build_limits}
\end{align}
\end{subequations}
Here, $\overline P_i$ denotes the rated electrical power of TCL $i$, and $\underline\vartheta_i$ and $\overline\vartheta_i$ define the prescribed indoor temperature range. The parameter $\alpha_i:=\exp(-T_sH_i/C_i)$ is the thermal retention factor, and $\mathrm{COP}_{i,k}$ is the heat pump coefficient of performance. The model assumes heating operation with a known initial indoor temperature $\vartheta_{i,0|t}$.

Following the thermal flexibility models in \cite{hao2015,zhao2017,li2023}, the baseline consumption profile is obtained by tracking the indoor-temperature setpoint. At each prediction step $k$, the lower and upper bounds on the individual energy deviation \(E_{i,k|t}\) are obtained by minimizing and maximizing the cumulative electrical-energy deviation from the baseline over the admissible power and temperature trajectories satisfying \eqref{eq:individual_build_envelope}. The corresponding aggregate power and energy deviations follow \eqref{eq:aggregate_power} and \eqref{eq:aggregate_energy}, respectively. For each TCL aggregation $r\in\mathcal D_{\textrm{BLDG}}$, the resulting bounds $(\underline E_{r,k|t},\overline E_{r,k|t})$ describe temporary load shifting and the subsequent rebound effect following load shifting relative to the baseline.

The EV and building load models above provide time-varying bounds on the power and cumulative energy deviation of each demand-side resource \cite{barot2017,wen2024}. Let $\mathbf x$ and $\mathbf u$ collect the response states and setpoints for ancillary service provisions over the prediction horizon. Similarly, $\mathbf p,\mathbf q\in\R^{N_fn_d}$ and $\mathbf E\in\R^{N_fn_{\mathrm{agg}}}$ collect the DER active power, reactive power, and aggregate energy deviations over the fast prediction horizon, where $n_{\mathrm{agg}}:=|\mathcal D_{\textrm{EV}}\cup\mathcal D_{\textrm{BLDG}}|\in\mathbb N$. The input and power sequences are indexed over $0:N_f-1$, and the response and energy states over $1:N_f$, with their initial values given. Combining the DER response models, capability constraints, and demand-side power-energy bounds defines the admissible power-energy set
\begin{equation}
\begin{aligned}
\mathcal X_t^{\textrm{PE}}:=\bigl\{(\mathbf{u},\mathbf{x},\mathbf{p},\mathbf{q},\mathbf{E}):\;&
\eqref{eq:device_model},\ \eqref{eq:physical_channel_changes},\ \eqref{eq:physical_output_composition},\eqref{eq:demand_energy},\\
&\eqref{eq:demand_energy_model},\
(\mathbf{p}_{k|t},\mathbf{q}_{k|t})\in\mathcal X^P_{k|t}\}.
\end{aligned}
\label{eq:power_energy_set}
\end{equation}
The predicted DER outputs contained in \(\mathcal X_t^{\mathrm{PE}}\) are subsequently used to evaluate the distribution system operating constraints and the ancillary-service delivery.

\subsection{Distribution System Model and Service Delivery}
We now consider a radial distribution feeder connected to the upstream grid at bus~0 and represented by the graph $\mathcal G=(\mathcal N_0,\mathcal E)$, where $\mathcal N_0$ is the bus set, $\mathcal N=\mathcal N_0\setminus\{0\}$ is the distribution bus set, and $\mathcal E\subseteq\mathcal N_0\times\mathcal N_0$ is the branch set, with each branch oriented away from bus~0. Their cardinalities $n_n=|\mathcal N|$ and $n_e=|\mathcal E|$ are positive integers. Let $\mathbf{s}_P\in\{-1,1\}^{n_d}$ encode the active power sign convention, with entries $+1$ for DGs, BESSs, and EV aggregations and $-1$ for aggregations of buildings with TCLs, and define $\mathbf{S}_P=\diag(\mathbf{s}_P)\in\R^{n_d\times n_d}$. The service-specific power changes are then mapped to net-injection changes as
\begin{subequations}\label{eq:service_net_changes}
\begin{align}
\Delta\mathbf p_{k|t}^{\PFC,\net}&=\mathbf S_P\Delta\mathbf p_{k|t}^{\PFC,\act},\label{eq:pfc_net_change}\\
\Delta\mathbf q_{k|t}^{\VC,\net}&=\Delta\mathbf q_{k|t}^{\VC,\act},\label{eq:vc_net_change}\\
\Delta\mathbf p_{k|t}^{\AGC,\net}&=\mathbf S_P\Delta\mathbf p_{k|t}^{\AGC,\act},\quad k\in\mathcal K.\label{eq:agc_net_change}
\end{align}
\end{subequations}
The total net-injection changes at each prediction step are
\begin{subequations}\label{eq:net_power_deviation}
\begin{align}
\Delta\mathbf p_{k|t}^{\net}&=\Delta\mathbf p_{k|t}^{\PFC,\net}+\Delta\mathbf p_{k|t}^{\AGC,\net},\label{eq:total_net_active_change}\\
\Delta\mathbf q_{k|t}^{\net}&=\Delta\mathbf q_{k|t}^{\VC,\net}.\label{eq:total_net_reactive_change}
\end{align}
\end{subequations}
These net-injection changes are then used to predict the corresponding bus voltage and branch current variations using a radial distribution system model linearized at the operating point. Let $\hat{\mathbf V}_t\in\C^{|\mathcal N|}$ and $\hat{\mathbf I}_t\in\C^{|\mathcal E|}$ collect the distribution bus voltages and branch currents at this operating point, and let $\Delta\mathbf v_{k|t}$ and $\Delta\mathbf i_{k|t}$ denote their predicted changes. 

To map the DER power variations to current-injection variations, define $\hat{\mathbf V}_{\mathcal D,t}\in\C^{|\mathcal D|}$ as the vector of bus voltages associated with the DERs. The resulting voltage-dependent mapping is
\begin{equation}
\mathbf{Y}_t=\diag\left((\hat{\mathbf{V}}_{\mathcal D,t})^*\right)^{-1}.
\label{eq:voltage_current_map}
\end{equation}

Then we use the reduced bus injection to branch current (BIBC) and direct load flow (DLF) matrices to obtain the predicted branch current and bus voltage changes as
\begin{subequations}\label{eq:network_sensitivity}
\begin{align}
\Delta\mathbf{i}_{k|t} & = -\mathbf{B}_R\mathbf{Y}_t\left(\Delta\mathbf{p}^{\net}_{k|t}-\mathrm{j}\,\Delta\mathbf{q}^{\net}_{k|t}\right), \label{eq:branch_current_sensitivity}\\
\Delta\mathbf{v}_{k|t} & = \mathbf{D}_R\mathbf{Y}_t\left(\Delta\mathbf{p}^{\net}_{k|t}-\mathrm{j}\,\Delta\mathbf{q}^{\net}_{k|t}\right), \label{eq:voltage_sensitivity}
\end{align}
\end{subequations}
where $\mathbf B_R\in\mathbb R^{|\mathcal E|\times|\mathcal D|}$ and $\mathbf D_R\in\mathbb C^{|\mathcal N|\times|\mathcal D|}$ denote the resource injection columns of the BIBC and DLF matrices, respectively \cite{teng2003}. Let $\mathbf F_t\in\C^{(n_n+n_e)\times2n_d}$ denote the resulting stacked voltage/current sensitivity matrix, evaluated at the current operating point and fixed over the prediction horizon. Combining these relations gives the compact linear sensitivity model

\begin{equation}
\begin{bmatrix}
\Delta\mathbf{v}_{k|t}\\
\Delta\mathbf{i}_{k|t}
\end{bmatrix}
=\mathbf{F}_t
\begin{bmatrix}
\Delta\mathbf{p}^{\net}_{k|t}\\
\Delta\mathbf{q}^{\net}_{k|t}
\end{bmatrix}.
\label{eq:compact_network_model}
\end{equation}

Since the distribution bus voltage angles remain close to the upstream reference angle, we approximate the voltage magnitudes by the real parts of the predicted complex voltages \cite{dhople2015}. We impose the voltage and branch current limits in a compact form:
\begin{equation}
\mathbf{g}_n\left(\hat{\mathbf{V}}_t+\Delta\mathbf{v}_{k|t},\hat{\mathbf{I}}_t+\Delta\mathbf{i}_{k|t}\right)\leq\mathbf{0},
\label{eq:network_security}
\end{equation}
where $\mathbf g_n:\C^{n_n}\times\C^{n_e}\to\R^{n_g^{\mathrm{net}}}$ collects the real-valued residuals of the voltage and branch current constraints, with $n_g^{\mathrm{net}}\in\mathbb N$. 

Let $\underline v\in\R$ and $\overline v\in\R$ denote the lower and upper limits for voltage magnitude. For each $e\in\mathcal E$, let $\mathbf A_e\in\R^{M_e\times2}$ and $\mathbf b_e\in\R^{M_e}$ specify a current limit polygon with $M_e\in\mathbb N$ facets. The scalar $\hat I_{e,t}$ and increment $\Delta i_{e,k|t}$ are the components associated with branch $e$:
\begin{subequations}\label{eq:network_limits}
\begin{gather}
\underline v\leq \Re\left(\hat{\mathbf V}_t+\Delta\mathbf v_{k|t}\right)
\leq\overline v,
\label{eq:voltage_limits}\\
\mathbf{A}_e
\begin{bmatrix}
\Im(\hat I_{e,t}+\Delta i_{e,k|t})\\
\Re(\hat I_{e,t}+\Delta i_{e,k|t})
\end{bmatrix}
\leq \mathbf{b}_e,\qquad e\in\mathcal E.
\label{eq:branch_current_limits}
\end{gather}
\end{subequations}
Both sets of constraints in~\eqref{eq:network_limits} are imposed at every $k\in\mathcal K$. Here, we approximate the magnitude limit of each branch current by a polygon.

The linearized distribution system model also determines each service's contribution to the power exchanged at the PCC. Let $\mathbf B_{R,1}\in\R^{1\times n_d}$ denote the row of the reduced BIBC matrix associated with the main feeder branch between the upstream bus and bus~1, and let $\hat V_{0,t}\in\C$ denote the upstream bus complex voltage. For each service $\nu\in\{\PFC,\VC,\AGC\}$ and prediction step $k\in\mathcal K$, let $\Delta\mathbf p_{k|t}^{\nu,\net}\in\R^{n_d}$ and $\Delta\mathbf q_{k|t}^{\nu,\net}\in\R^{n_d}$ denote its predicted changes in net active and reactive power injections in the common DER order, with unused components set to zero. The resulting main feeder complex power change is
\begin{equation}
\label{eq:physical_delivery}
\begin{aligned}
\Delta S_{0,k|t}^{\nu}=&\hat V_{0,t}\Bigl[\mathbf B_{R,1}\mathbf Y_t
\bigl(\Delta\mathbf p_{k|t}^{\nu,\net}
-\mathrm j\Delta\mathbf q_{k|t}^{\nu,\net}\bigr)\Bigr]^*,
\end{aligned}
\end{equation}
where $\Delta S_{0,k|t}^{\nu}\in\C$ is the complex power increment. Let $\hat y_t^\nu\in\R$ denote the service contribution at the PCC before the current update. Adding the predicted change gives the total service deliveries: 
\begin{subequations}\label{eq:service_total_delivery}
\begin{align}
y_{k|t}^{\PFC}&=\hat y_t^{\PFC}+\Re\{\Delta S_{0,k|t}^{\PFC}\},\\
y_{k|t}^{\VC}&=\hat y_t^{\VC}+\Im\{\Delta S_{0,k|t}^{\VC}\},\\
y_{k|t}^{\AGC}&=\hat y_t^{\AGC}+\Re\{\Delta S_{0,k|t}^{\AGC}\}.
\end{align}
\end{subequations}
The predicted service deliveries are used to formulate the service delivery constraints in the joint MPC problem.

\subsection{Joint MPC Formulation for Ancillary Services}
\label{sec:mpc}
Combining the DER response, demand-side flexibility, and distribution system models, we obtain discrete-time representations of the dynamics in~\eqref{eq:general_ct_dynamics}, equality constraints in~\eqref{eq:general_ct_eq}, and inequality constraints in~\eqref{eq:general_ct_ineq}. Accordingly, we formulate a finite-horizon optimization problem to jointly coordinate PFC, VC, and AGC by optimizing the service setpoint increments and service delivery mismatches. The decision variables are collected as
\begin{equation}
\label{eq:decision_trajectory}
\begin{aligned}
\bm\Omega_t^c:=\bigl(&
(\Delta\mathbf u_{k|t}^{\PFC},\eta_{k|t}^{\PFC})_{k\in\mathcal K},\\
&(\Delta\mathbf u_{k|t}^{\VC},\eta_{k|t}^{\VC})_{k\in\mathcal K},\\
&(\Delta\mathbf u_{k|t}^{\AGC},\eta_{k|t}^{\AGC})_{k\in\mathcal A_m}\bigr).
\end{aligned}
\end{equation}
where the signed slack variables $\eta^{\PFC}_{k|t}$, $\eta^{\VC}_{k|t}$, and $\eta^{\AGC}_{k|t}$ quantify the mismatch between the predicted delivery of each service at the PCC and the request. Exact delivery of all service requests may not be feasible because the services share the same DER capability, network operating limits, and demand-side power-energy flexibility. These mismatch variables therefore relax the service-delivery equalities when the requested support cannot be fully provided.

Let $\mathbf w_P$, $\mathbf w_V$, and $\mathbf w_A$ be nonnegative resource weight vectors in $\R_{\geq0}^{n_d}$, and define $\mathbf W_{\PFC}=\diag(\mathbf w_P)$, $\mathbf W_{\VC}=\diag(\mathbf w_V)$, and $\mathbf W_{\AGC}=\diag(\mathbf w_A)$. These matrices belong to $\R^{n_d\times n_d}$ and are positive semidefinite. Let $\gamma>0$ be the service mismatch penalty. The first three objective terms penalize PFC, VC, and AGC updates, and the last term penalizes service delivery mismatches:
\begin{equation}\label{eq:central_objective}
\begin{aligned}
J_t^c=&\sum_{k\in\mathcal K}\|\Delta\mathbf u_{k|t}^{\PFC}\|_{\mathbf W_{\PFC}}^2
+\sum_{k\in\mathcal K}\|\Delta\mathbf u_{k|t}^{\VC}\|_{\mathbf W_{\VC}}^2\\
&+\sum_{k\in\mathcal A_m}\|\Delta\mathbf u_{k|t}^{\AGC}\|_{\mathbf W_{\AGC}}^2\\
&+\gamma\left(\sum_{k\in\mathcal K}|\eta_{k|t}^{\PFC}|
+\sum_{k\in\mathcal K}|\eta_{k|t}^{\VC}|
+\sum_{k\in\mathcal A_m}|\eta_{k|t}^{\AGC}|\right).
\end{aligned}
\end{equation}
Together with the prediction models and constraints, this objective defines the centralized multi-rate MPC problem:
\begin{subequations}\label{eq:central_mpc}
\begin{align}
\hspace{-5mm}\bm{\mathcal P}_t^c:\;\underset{\bm\Omega_t^c}{\min}\;\;\;
&J_t^c(\bm\Omega_t^c),\label{eq:central_problem}\\
\hspace{-5mm}\text{subject to}\;\;\;
&(\mathbf u,\mathbf x,\mathbf p,\mathbf q,\mathbf E)
\in\mathcal X_t^{\mathrm{PE}},\label{eq:central_device_constraints}\\
\hspace{-5mm}&\begin{bmatrix}\Delta\mathbf v_{k|t}\\\Delta\mathbf i_{k|t}\end{bmatrix}
=\mathbf F_t\begin{bmatrix}\Delta\mathbf p_{k|t}^{\net}\\
\Delta\mathbf q_{k|t}^{\net}\end{bmatrix},
\; k\in\mathcal K,\label{eq:central_network_model}\\
\hspace{-5mm}&\mathbf g_n\bigl(\hat{\mathbf V}_t+\Delta\mathbf v_{k|t},
\hat{\mathbf I}_t+\Delta\mathbf i_{k|t}\bigr)\leq\mathbf0,
\; k\in\mathcal K,\label{eq:central_network_constraints}\\
\hspace{-5mm}&y_{k|t}^{\PFC}=\rho_t^{\PFC}+\eta_{k|t}^{\PFC},
\quad k\in\mathcal K,\label{eq:central_pfc_matching}\\
\hspace{-5mm}&y_{k|t}^{\VC}=\rho_t^{\VC}+\eta_{k|t}^{\VC},
\quad k\in\mathcal K,\label{eq:central_vc_matching}\\
\hspace{-5mm}&y_{k|t}^{\AGC}
=\rho_t^{\AGC}+\eta_{k|t}^{\AGC},
\quad k\in\mathcal A_m.\label{eq:central_agc_matching}
\end{align}
\end{subequations}
Constraint~\eqref{eq:central_device_constraints} collects the DER dynamics, capability limits, and the power--energy flexibility envelopes of the EV and TCL aggregations. The distribution system model in \eqref{eq:central_network_model} maps predicted net injection changes to voltage and branch current changes, and \eqref{eq:central_network_constraints} imposes limits on the resulting voltages and currents. Constraints~\eqref{eq:central_pfc_matching}--\eqref{eq:central_agc_matching} match the total predicted PFC, VC, and AGC deliveries to their respective service requests, with slack variables allowing for service mismatches. The signed slack variables relax only the service delivery equalities; the DER and network operating constraints remain hard constraints.

To implement C-MPC in receding horizon form, we solve problem~\eqref{eq:central_mpc} over the shifted horizon at each fast update using the updated operating point and states. Algorithm~\ref{alg:mpc} summarizes this receding horizon procedure.
\begin{algorithm}[!htbp]
\caption{Receding horizon multi-rate coordination}\label{alg:mpc}
\begin{algorithmic}[1]
\State Initialize the operating point, DER setpoints, and prediction model states.
\For{each fast update instant $t_m=mT_{\mathrm{MPC}}$}
    \State Acquire feeder and DER measurements; update the operating point, $\hat{\mathbf x}_{t_m}^{\nu}$, and $\hat E_{r,t_m}$.
    \State Update the distribution system model at the current operating point and compute the errors using \eqref{eq:service_errors}.
    \State Calculate the PFC and VC requests using \eqref{eq:pfc_request} and \eqref{eq:vc_request}, and sample the AGC request using \eqref{eq:agc_request_sampling}.
    \State Express the setpoint trajectories $\mathbf u^{\nu}$, response states $\mathbf x^{\nu}$, and outputs $\mathbf r^{\nu}$ in terms of the setpoint increments using \eqref{eq:general_increment} and \eqref{eq:device_model}.
    \State Express the physical outputs $\mathbf p$ and $\mathbf q$ using \eqref{eq:physical_channel_changes}--\eqref{eq:physical_output_composition}; form the aggregate energy deviation $\mathbf E$ using \eqref{eq:demand_energy}.
    \State Impose the DER capability limits in \eqref{eq:instantaneous_capability_set} and the power--energy flexibility envelopes in \eqref{eq:demand_energy_model} to form $\mathcal X_{t_m}^{\textrm{PE}}$ defined in \eqref{eq:power_energy_set}.
    \State Construct and solve problem~\eqref{eq:central_mpc} at $t=t_m$ with the setpoint update constraints in \eqref{eq:general_update_constraint} to obtain $\bm\Omega_{t_m}^{c,\star}$.
    \State Apply the first optimized PFC and VC setpoints $\mathbf u_{0|t_m}^{\PFC,\star}$ and $\mathbf u_{0|t_m}^{\VC,\star}$.
    \If{$m\bmod q=0$}
        \State Apply the first optimized AGC setpoint $\mathbf u_{0|t_m}^{\AGC,\star}$.
    \Else
        \State Retain the previously applied AGC setpoint.
    \EndIf
    \EndFor
\end{algorithmic}
\end{algorithm}
Each optimization updates the service allocation over the prediction horizon using the current measurements and service requests. The first optimized PFC and VC setpoints are applied at every fast update, whereas the AGC setpoint is updated only when $m\bmod q=0$. These setpoints are jointly optimized using the DER response models, with the DER capability limits, power--energy envelopes, and distribution system constraints imposed on the combined physical response.

\section{Numerical Evaluation}
\label{sec:results}
\begin{figure}[!htbp]
\centering
\includegraphics[width=\columnwidth]{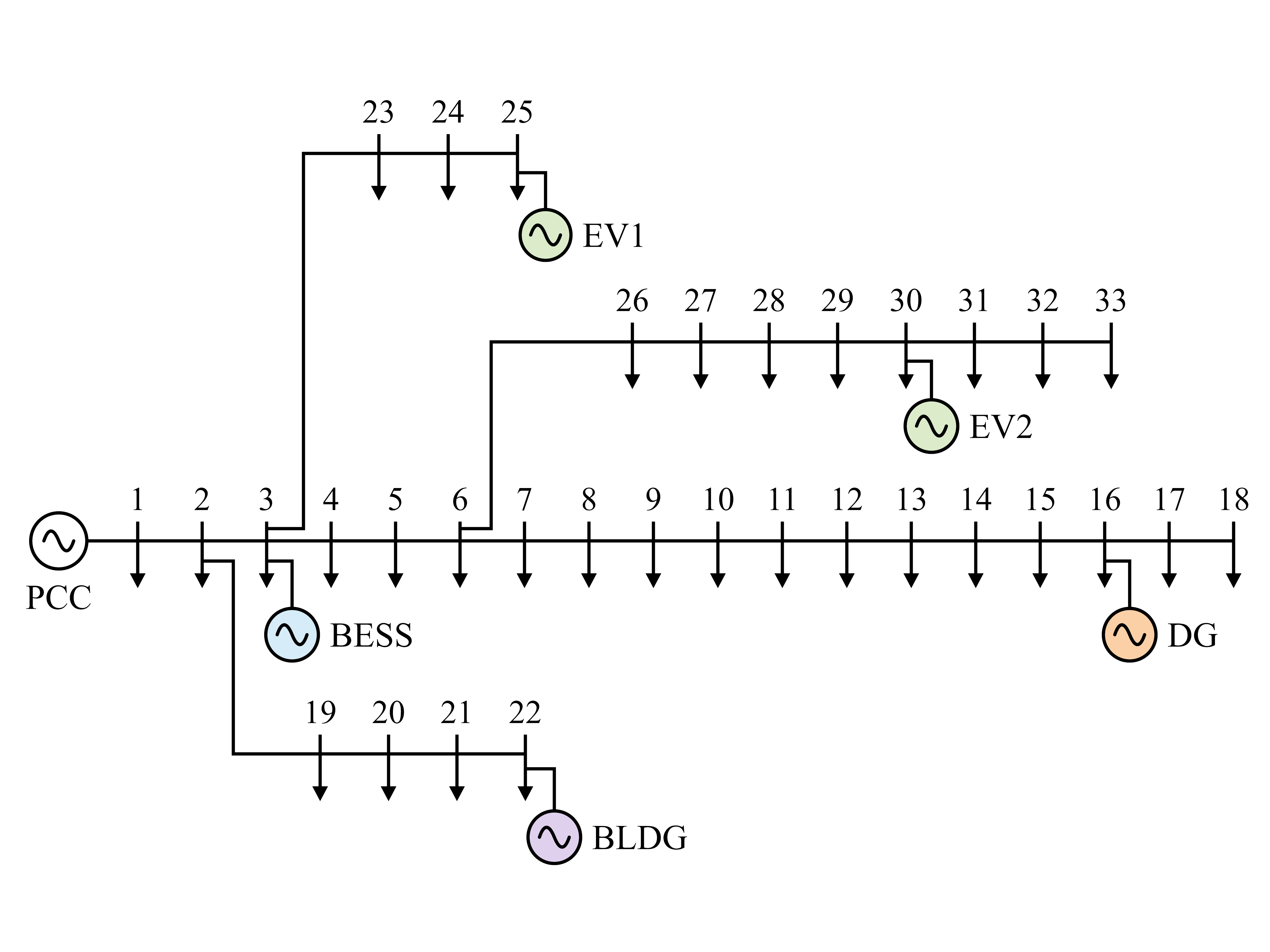}
\caption{Modified IEEE 33-bus system with controllable resources.}
\label{fig:topology}
\end{figure}

In this section, we evaluate how the coordination architecture and service update rates affect control performance using a modified IEEE 33-bus DN based on the radial feeder in~\cite{baran1989}. Fig.~\ref{fig:topology} shows the network connection to the upstream grid and the locations of the DERs: a BESS at bus~3, a diesel generator at bus~16, two EV aggregations at buses~25 and~30, and a TCL aggregation at bus~22. Both EV aggregations support vehicle-to-grid (V2G) operation through bidirectional converters, allowing active power discharge and reactive power support when the vehicles are connected.

For validation under closed-loop control, we use a nonlinear differential-algebraic equation (DAE) model of the DN and its connected resources. The model includes the upstream frequency equivalent, diesel generator and governor dynamics, converter and filter dynamics of the BESS and EV aggregations, and building thermal dynamics. Dynamic branch currents are coupled with these component models through algebraic network equations. These component and distribution system models follow standard power system DAE formulations~\cite{kundur1994,milano2010}; related dynamic representations of active DNs are given in~\cite{chaspierre2020,fulgencio2020}. We use this model to simulate the physical system response and evaluate voltage, current, frequency, and cumulative energy deviations.

We set $T_s=T_{sh}=T_{\mathrm{MPC}}=T_f=1$~s and $N_f=20$, giving a prediction horizon of 20~s. Multi-rate control updates PFC and VC every 1~s and AGC every 10~s. Uniformly fast control updates all three services every 1~s, and uniformly slow control updates them every 10~s. The AGC request is sampled every 1~s, independently of these setpoint execution periods. An external AGC request is generated offline from a previously recorded reference frequency response and supplied as an input to the controllers.

The power base is 1~MVA and the voltage base is 12.66~kV. To maintain voltage security, bus voltages are constrained to $[0.95,1.05]$~p.u. Frequency indices are calculated from the nonlinear DAE simulations, and control effort is evaluated from the setpoint increments applied at the corresponding service updates. 

To assess frequency regulation and control effort, we use the integral absolute error (IAE) of the frequency deviation and the index $J_{\textrm{move}}$, respectively. The IAE is defined as
\begin{equation}
J_f=\int_{t_0}^{t_1}|f(t)-f_0|\,\textrm{d}t,
\label{eq:frequency_iae}
\end{equation}
where $f(t)$ is the simulated system frequency, $f_0=50$~Hz is the reference frequency, and $t_0$ and $t_1$ denote the disturbance onset and the end of the simulation. The frequency deviation is defined as $\Delta f(t):=f(t)-f_0$. To quantitatively evaluate the frequency recovery performance, we define the settling time as the elapsed time from the event to the first instant after which $|f(t)-f_0|\leq0.01$~Hz holds until the end of the simulation.

Let $\Delta\mathbf u^{\PFC}_m$, $\Delta\mathbf u^{\VC}_m$, and $\Delta\mathbf u^{\AGC}_m\in\R^{n_d}$ denote the setpoint increments actually applied at update instant $t_m$. The control effort is quantified by 
\begin{align}\notag
J_{\textrm{move}}=&\frac{1}{T_{\PFC}}\sum_{m\in\mathcal M_{\PFC}}
\|\Delta\mathbf{u}_m^{\PFC}\|_{\mathbf{W}_{\PFC}}^2
+\frac{1}{T_{\VC}}\sum_{m\in\mathcal M_{\VC}}
\|\Delta\mathbf{u}_m^{\VC}\|_{\mathbf{W}_{\VC}}^2\\
&+\frac{1}{T_a}\sum_{m\in\mathcal M_{\AGC}}
\|\Delta\mathbf{u}_m^{\AGC}\|_{\mathbf{W}_{\AGC}}^2,
\label{eq:move_effort}
\end{align}
where $\mathcal M_{\PFC}$, $\mathcal M_{\VC}$, and $\mathcal M_{\AGC}$ denote the actual update indices for PFC, VC, and AGC during the 150~s following the event, with corresponding update periods $T_{\PFC}$, $T_{\VC}$, and $T_a$. Each service contribution to $J_{\textrm{move}}$ is obtained by summing its weighted squared setpoint increments over these updates and dividing the sum by its update period. With power expressed in per unit and dimensionless weights, $J_{\textrm{move}}$ is reported in p.u.$^2$/s. 

We perform the simulations in MATLAB R2023b, model the MPC problems in YALMIP \cite{lofberg2004}, and solve them using COPT with its default numerical parameters on a workstation with an Intel Core Ultra X7 358H processor and 32~GB RAM.

Three case studies are conducted to evaluate different aspects of controller performance. Case~1 compares C-MPC and hierarchical model predictive control (H-MPC) under a generation loss event. Case~2 uses the same disturbance to evaluate the effect of service update rates, while Case~3 examines voltage support following an upstream line trip.

\subsection{Centralized Versus Hierarchical Coordination}
We first compare C-MPC and H-MPC under the generation loss event in Case~1. During the first 300~s, the system is allowed to settle to a steady operating condition. At $t=300$~s, a loss of generation creates an active power deficit of 0.40~p.u. The responses of both controllers are compared over the following 150~s.

\begin{figure*}[!htbp]
\centering
\includegraphics[width=0.95\textwidth]{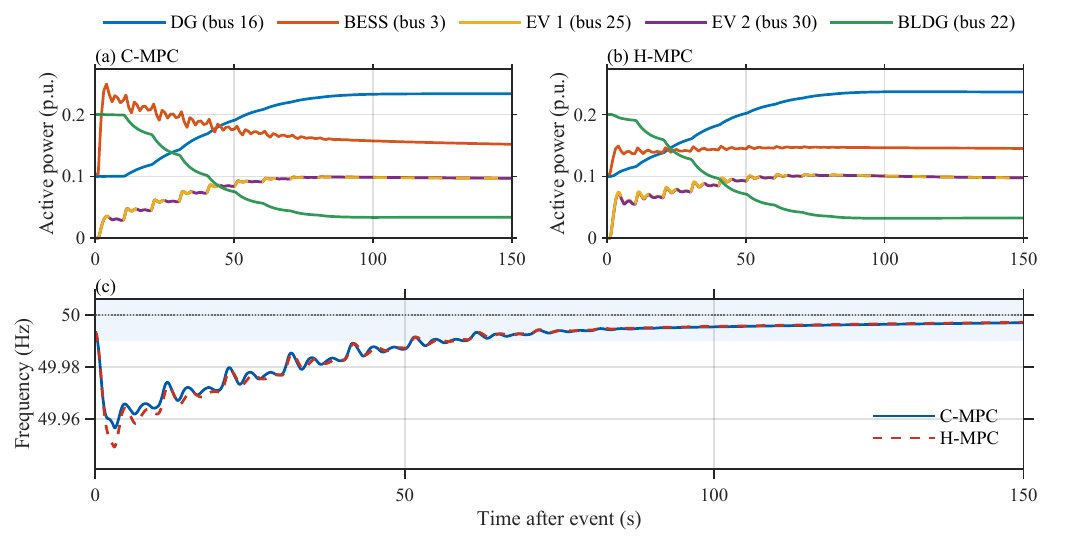}
\caption{Case~1 responses to the active power deficit: (a) DER active power outputs under C-MPC; (b) DER active power outputs under H-MPC; (c) system frequency.}
\label{fig:case1active}
\end{figure*}

As shown in Fig.~\ref{fig:case1active}(c), C-MPC limits the initial frequency decline more effectively than H-MPC, while the subsequent frequency responses remain close. The frequency nadir is 49.9567~Hz under C-MPC and 49.9493~Hz under H-MPC. The corresponding IAEs are 1.7230~Hz$\cdot$s for C-MPC and 1.7649~Hz$\cdot$s for H-MPC, with settling times of 57.42~s and 57.90~s, respectively. Although the two architectures provide comparable frequency regulation, they allocate the available DER flexibility differently.

The active power responses in Fig.~\ref{fig:case1active}(a) and (b) show how the two controllers achieve similar frequency regulation with different DER allocations. Immediately after the event, the BESS output rises rapidly, whereas the DG contribution develops over several tens of seconds. With C-MPC, the BESS output increases by up to 0.145~p.u. relative to its pre-event value and then decreases as the other DER outputs grow. By comparison, H-MPC yields a maximum increase of 0.045~p.u., with the BESS output remaining near that level during the subsequent response. Both EV aggregations increase their active power outputs in both cases. Despite their different feeder locations, the two EV aggregations exhibit similar responses, consistent with their identical parameter settings. TCL power consumption decreases gradually to approximately 0.03~p.u. and remains near that level. A negative setpoint to the TCL aggregation represents this reduction in consumption and contributes positive support at the PCC. 

This difference results from the different coordination strategies used by the two controllers. Specifically, C-MPC jointly optimizes the PFC and AGC setpoints using predicted DER outputs, allowing the BESS to provide rapid initial support and gradually reduce its contribution as the other DERs increase their support. In contrast, H-MPC holds the AGC allocation between upper-layer updates, while its fast layer adjusts the PFC setpoints through droop control based on the measured frequency error. This separate allocation does not explicitly coordinate the PFC setpoints with the developing AGC response, resulting in a less pronounced transfer of support among the DERs.

\begin{figure*}[!htbp]
\centering
\includegraphics[width=0.95\textwidth]{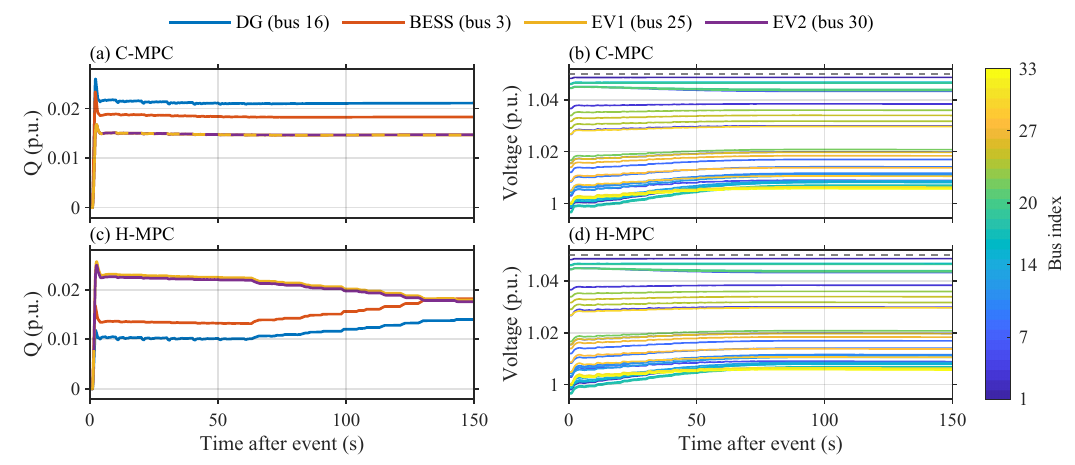}
\caption{Case~1 reactive power and voltage responses: (a), (b) C-MPC; (c), (d) H-MPC. Left: DER reactive power outputs. Right: bus voltages.}
\label{fig:case1reactive}
\end{figure*}

The active power redispatch also changes feeder flows and bus voltages, requiring reactive power adjustments. Accordingly, the reactive power allocations in Fig.~\ref{fig:case1reactive}(a) and (c) show a more pronounced difference between the two architectures than the voltage responses alone. C-MPC assigns larger initial reactive power contributions to the DG and BESS, with smaller and nearly constant contributions from the two EV aggregations. H-MPC initially obtains more reactive power support from the EV aggregations and less from the DG and BESS. When EV departures begin at about 60~s, the EV reactive power outputs under H-MPC decrease as the DG and BESS outputs increase. Under C-MPC, the reactive power outputs of these resources change much less.

The initial reactive power allocation to the EV aggregations under C-MPC remains within their available capability as vehicles depart. Since the objective in \eqref{eq:central_objective} penalizes VC setpoint changes, subsequent optimizations maintain nearly the same EV reactive power outputs. Under H-MPC, the VC participation factors are proportional to the available capacities and are updated as EV availability decreases. Consequently, the EV share decreases and the DG and BESS shares increase, leading to the redistribution shown in Fig.~\ref{fig:case1reactive}(c). Despite these different allocations, the bus voltage profiles in Fig.~\ref{fig:case1reactive}(b) and (d) remain close. Increased DER active power injection, reduced TCL power consumption, and reactive power support accompany a gradual voltage rise at many buses. Bus voltages under both controllers remain within $[0.9965,1.0487]$~p.u. 

\begin{figure}[!htbp]
\centering
\includegraphics[width=0.68\columnwidth]{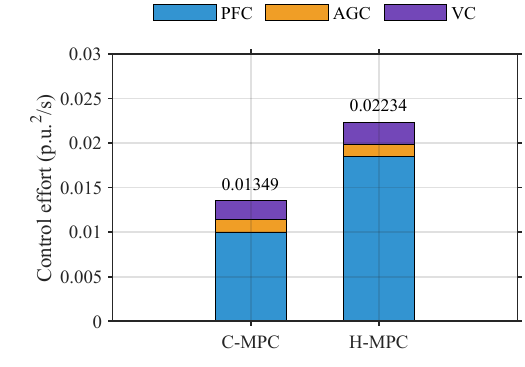}
\caption{Case 1 comparison of total control effort and its PFC, VC, and AGC contributions under C-MPC and H-MPC.}
\label{fig:case1summary}
\end{figure}

Finally, we compare the control effort results in Fig.~\ref{fig:case1summary}, which distinguish the two architectures despite their similar frequency and voltage responses. PFC accounts for most of the difference: its contribution to the control effort is 45.92\% lower under C-MPC, and the AGC contributions remain close with the same update period. The total control effort is 0.01349 for C-MPC and 0.02234 for H-MPC, a reduction of 39.60\%. The larger PFC contribution to the control effort under H-MPC is consistent with its fast layer continuing to adjust the droop setpoints as the frequency error changes between upper layer AGC updates. In C-MPC, the PFC and VC setpoints are jointly optimized with the AGC setpoint while accounting for the developing AGC response. This coordination reduces repeated PFC corrections without noticeably changing frequency recovery.

\subsection{Effect of Service Update Rates}
\begin{figure*}[!htbp]
\centering
\includegraphics[width=1.00\textwidth]{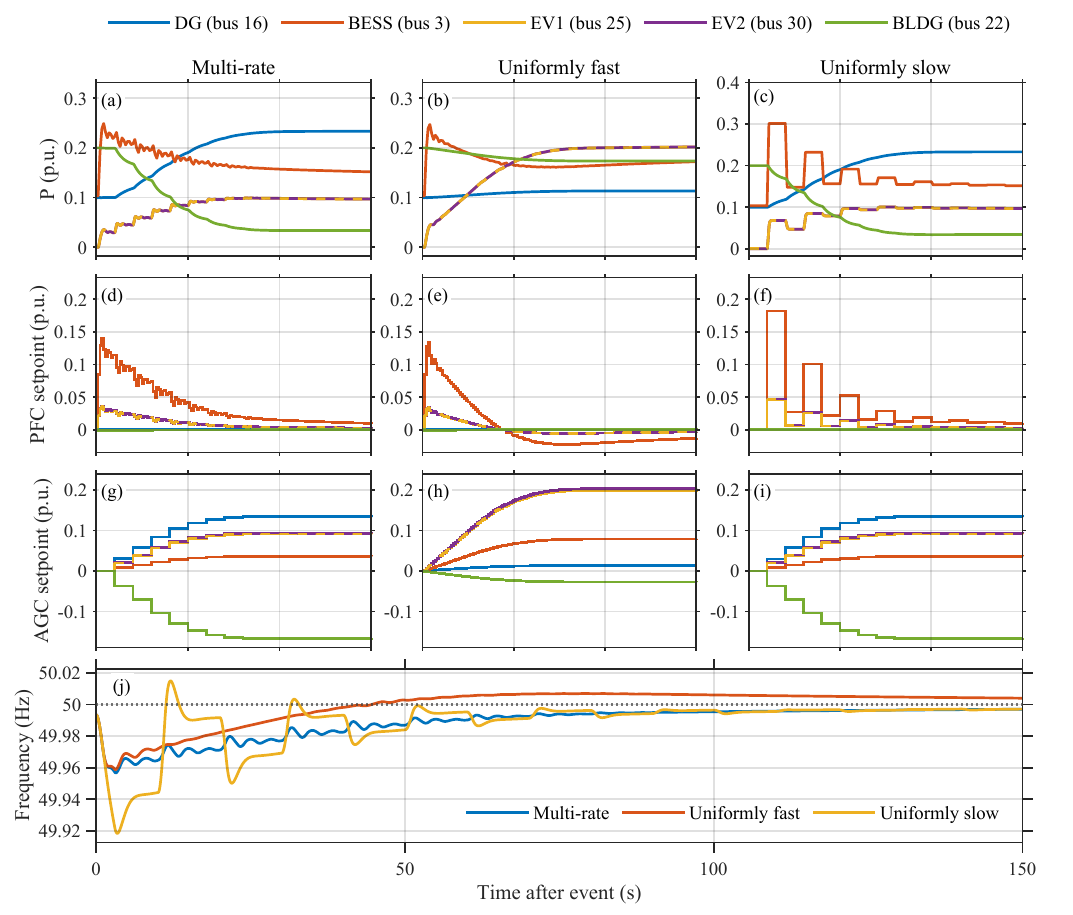}
\caption{Case~2 responses under different update schedules: (a)--(c) DER active power outputs; (d)--(f) PFC setpoints; (g)--(i) AGC setpoints; (j) system frequency.}
\label{fig:rate}
\end{figure*}

In Case~2, we compare the multi-rate, uniformly fast, and uniformly slow implementations of C-MPC under the same active power deficit event as in Case~1. The comparison examines how service update rates affect frequency regulation, DER allocation, and computational requirements. The frequency responses in Fig.~\ref{fig:rate}(j) first differ during the initial decline. Multi-rate control updates PFC every second and limits the maximum frequency deviation to 0.0433~Hz, close to the 0.0412~Hz obtained under uniformly fast control. Both schedules therefore provide prompt initial regulation through the fast resources. Their subsequent responses depend on how AGC provides sustained support and how the PFC allocation changes as the other DER outputs develop.

Under multi-rate control, the BESS and EV aggregations respond promptly after the event, while the DG output increases and TCL power consumption decreases more gradually, as shown in Fig.~\ref{fig:rate}(a). The corresponding PFC and AGC setpoints in panels (d) and (g) show how the services share the available DER flexibility during this process. The initial PFC support is concentrated on the faster resources, and the DG and TCL aggregation subsequently provide a substantial share of the sustained support. The joint optimization accounts for these developing physical outputs when revising the service allocation. Thus, fast regulation can be adjusted as the slower resources increase their contributions, rather than requiring the initial support to be maintained by the same DERs throughout the response.

During the subsequent recovery, uniformly fast control begins to differ more clearly from multi-rate control. Its AGC setpoints increase at successive 1~s updates in Fig.~\ref{fig:rate}(h), compared with the 10~s updates in panel (g). These frequent adjustments assign larger sustained contributions to the BESS and EV aggregations, while the increase in DG output and the reduction in TCL power consumption are smaller in panel (b). Since the external AGC request, device limits, and resource weights are unchanged, the different allocations reflect the effect of the service update schedule. In this case, uniformly fast control relies more heavily on the faster resources to provide sustained support, whereas multi-rate control makes greater use of the available flexibility of the DG and TCL aggregation.

The larger sustained contributions from the fast resources also accelerate frequency recovery. Uniformly fast control has a settling time of 28.59~s and a frequency IAE of 1.2774~Hz$\cdot$s, compared with 57.42~s and 1.7230~Hz$\cdot$s for multi-rate control. The frequency subsequently exceeds 50~Hz, reaching a maximum of 50.0069~Hz. During this stage, the BESS and EV PFC setpoints become negative in Fig.~\ref{fig:rate}(e), providing regulation opposite to the sustained AGC support. These negative PFC contributions are part of the combined service allocation and do not imply negative total DER active power outputs. Multi-rate control produces a more gradual recovery while distributing more of the sustained support to the slower resources. Despite these different allocations, the two schedules have similar control effort, with values of 0.01349 under multi-rate control and 0.01299 under uniformly fast control.

The importance of preserving fast regulation becomes clearer in the comparison with uniformly slow control. Updating all service setpoints every 10~s delays the initial PFC correction, allowing the frequency to fall to 49.9184~Hz. The resulting maximum frequency deviation is 0.0816~Hz, substantially larger than under the other two schedules. The larger PFC adjustments in Fig.~\ref{fig:rate}(f) then produce a rapid recovery immediately after the nadir, followed by more pronounced oscillations. The frequency IAE is 1.7669~Hz$\cdot$s and the settling time is 64.53~s. Although its AGC allocation makes similar use of the DG and TCL aggregation, uniformly slow control cannot combine this sustained support with the prompt initial regulation obtained under multi-rate control.

These response differences also need to be considered together with the computational requirements. Updating AGC at every fast prediction step introduces more independent AGC decisions into the finite-horizon problem. Multi-rate control avoids these additional decisions while retaining fast PFC and VC updates. Table~\ref{tab:rate} reports a mean COPT solve time of 0.0731~s for multi-rate control, compared with 0.1009~s for uniformly fast control, a reduction of approximately 27.6\%. Uniformly slow control has the shortest mean COPT solve time of 0.0394~s, but its less frequent PFC and VC updates lead to a larger initial frequency deviation. Multi-rate control therefore preserves prompt initial regulation with a smaller online optimization problem and a lower mean COPT solve time than uniformly fast control.

\begin{table}[!htbp]
\caption{Case~2 Performance Across Update Rates and Mean COPT Solve Time}\label{tab:rate}
\centering
\renewcommand{\arraystretch}{1.2}
\footnotesize
\setlength{\tabcolsep}{3.5pt}
\begin{tabular}{lcccc}
\hline
Schedule & \shortstack{IAE\\(Hz$\cdot$s)} & \shortstack{Max. $|\Delta f|$\\(Hz)} & \shortstack{Control Effort\\(p.u.$^2$/s)} & \shortstack{Mean COPT\\solve time (s)}\\
\hline
Multi-rate & 1.7230 & 0.0433 & 0.01349 & 0.0731\\
Uniformly fast & 1.2774 & 0.0412 & 0.01299 & 0.1009\\
Uniformly slow & 1.7669 & 0.0816 & 0.00713 & 0.0394\\
\hline
\end{tabular}
\end{table}

For larger DNs, the number of service setpoints increases with the number of participating DERs, making the size of the online optimization increasingly relevant to implementation within a fast control period. The multi-rate structure limits the additional AGC decisions without slowing the updates required for PFC and VC. This reduction may help accommodate more DERs within the available computation time, while their different physical responses remain coordinated over a common prediction horizon. In this setting, accepting a longer subsequent frequency recovery can be reasonable when prompt initial regulation, the use of slower DER flexibility, and online computational requirements need to be addressed together. The results therefore support multi-rate coordination as a practical choice for this application, rather than updating all services at either the fast or the slow rate. All three schedules satisfy the aggregate demand-side power--energy flexibility envelopes, with no solver failures, voltage limit violations, or branch current overloads.

\subsection{Sensitivity to MPC Timing Parameters}
\begin{figure}[!htbp]
\centering
\includegraphics[width=0.88\columnwidth]{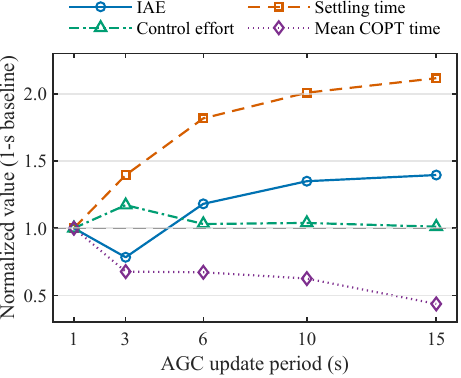}
\caption{Case~2 performance metrics for different AGC update periods. Each metric is normalized by its value with 1~s AGC updates.}
\label{fig:agcperiod}
\end{figure}

We further examine the effect of the AGC update period while maintaining fast PFC and VC updates. The AGC setpoints are updated every 1, 3, 6, 10, or 15~s, while the PFC and VC update periods, MPC execution period, and AGC request sampling period remain at 1~s. For this comparison, the prediction horizon is increased to 30~s for all five implementations, since 30~s is an integer multiple of the five AGC update periods. Fig.~\ref{fig:agcperiod} compares the frequency IAE, settling time, control effort, and mean COPT solve time.

The settling time increases from 28.59 to 60.50~s as the AGC update period increases from 1 to 15~s. This slower recovery is consistent with the less frequent adjustment of the AGC setpoints that provide sustained support. The frequency IAE, however, reaches its minimum of 1.0010~Hz$\cdot$s with 3~s AGC updates, which is 21.6\% below the 1~s value, while the settling time increases to 39.89~s. Under 1~s AGC updates, the faster frequency recovery is followed by an overshoot, which also contributes to the accumulated frequency deviation. Thus, the shortest settling time does not coincide with the lowest IAE in this case. 

The control effort also varies with the AGC update period. Its largest value of 0.01522 occurs with 3~s updates and is approximately 17.2\% higher than the 1~s value. The lowest frequency IAE is therefore obtained with greater control effort in this comparison. With 6, 10, and 15~s updates, the control effort remains within 4\% of the 1~s value. These results show that less frequent AGC updates do not necessarily reduce the overall control effort, since the index accounts for the adjustments of all three services as their joint allocation changes.

These response differences are accompanied by changes in computational requirements. Over the fixed prediction horizon, longer AGC update periods reduce the number of independent AGC decisions. The recorded mean COPT solve time decreases from 0.1296~s with 1~s updates to 0.0566~s with 15~s updates. Therefore, the AGC update period can be selected according to the regulation and computational requirements. The multi-rate structure maintains the fast updates required for PFC and VC while accommodating slower AGC updates. The proposed framework thus coordinates fast regulation and sustained support through a joint allocation of available DER flexibility across different time scales.

\begin{figure}[!htbp]
\centering
\includegraphics[width=0.98\columnwidth]{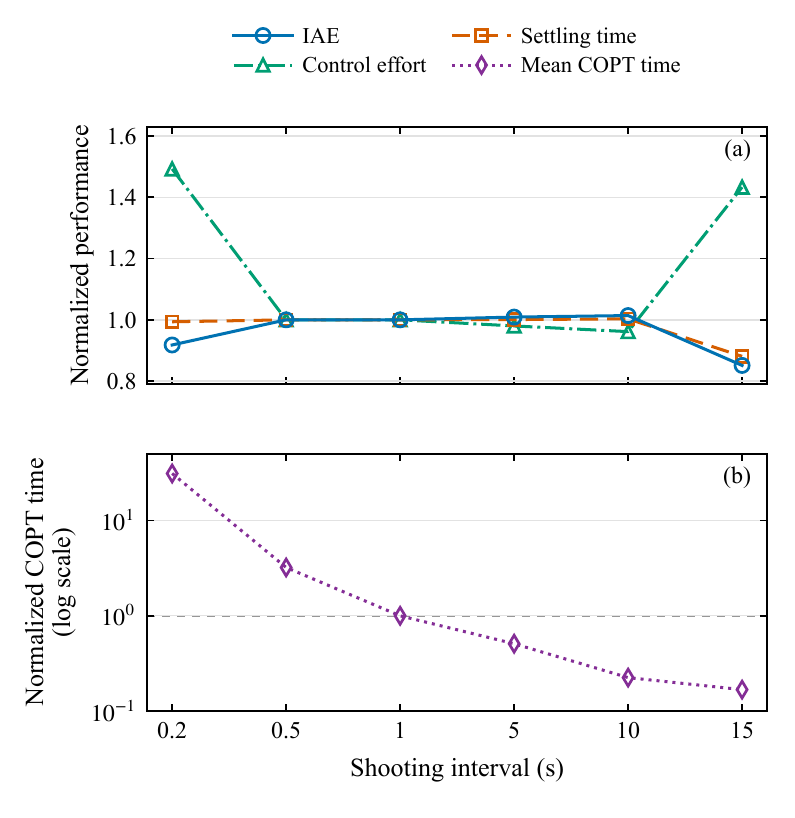}
\caption{Sensitivity to the shooting interval: (a) frequency IAE, settling time, and control effort; (b) mean COPT solve time on a logarithmic scale. Each metric is normalized by its value with $T_{\mathrm{sh}}=1$~s.}
\label{fig:shooting}
\end{figure}

We next examine the effect of the shooting interval with the service update periods fixed. The prediction horizon remains at 30~s, and $T_{\mathrm{sh}}$ is set to 0.2, 0.5, 1, 5, 10, or 15~s. At each 1~s MPC update, the controller uses the selected shooting interval to predict the DER responses and evaluate the operating constraints. Fig.~\ref{fig:shooting} compares the resulting frequency indices, control effort, and mean COPT solve time.

As shown in Fig.~\ref{fig:shooting}(a), the frequency indices and control effort remain close with 0.5 and 1~s shooting intervals. Increasing the interval to 5 or 10~s also has a relatively small effect: the IAE remains within 1.4\% of the 1~s value, and the settling time increases by less than 0.2~s. With a 0.2~s shooting interval, the IAE decreases from 1.7240 to 1.5827~Hz$\cdot$s, while the settling time decreases from 57.42 to 57.06~s. This reduction in frequency deviation is accompanied by greater control effort, which increases from 0.01351 to 0.02016, or by 49.2\%, mainly through larger PFC adjustments.

With a 15~s shooting interval, the frequency IAE decreases to 1.4677~Hz$\cdot$s and the settling time to 50.58~s. However, the control effort increases to 0.01935, which is 43.3\% higher than the 1~s value. As in the preceding comparison of AGC update periods, the improved frequency regulation is therefore obtained with greater control effort. These results also show that the frequency indices do not vary monotonically with the shooting interval.

In contrast, the mean COPT solve time in Fig.~\ref{fig:shooting}(b) decreases consistently as the shooting interval increases. Over the fixed prediction horizon, larger shooting intervals reduce the number of predicted states and independent setpoint decisions. Increasing the interval from 0.5 to 1~s reduces the mean COPT solve time from 0.4621 to 0.1439~s with little change in the frequency indices or control effort. With 5 and 10~s shooting intervals, the mean COPT solve time is further reduced by 49.0\% and 77.5\%, respectively, relative to the 1~s value. By comparison, the 0.2~s implementation requires a mean COPT solve time of 4.4627~s, exceeding the 1~s MPC execution period. Thus, the shooting interval can be selected according to the regulation and computational requirements while maintaining the prescribed service update periods.

\subsection{Voltage Support Under a Line Trip Event}
We then consider voltage support following an upstream line trip in Case~3. We represent this event by increasing the equivalent resistance of the upstream grid by 0.005~p.u. in the reduced Th\'evenin model. This step change raises the resistance from 0.0004 to 0.0054~p.u., which is 13.5 times its value before the event. PFC and AGC are disabled, and the upstream source voltage setpoint and active power setpoints remain unchanged. Hence, the response primarily reflects reactive power support under the changed network condition.

\begin{figure}[!htbp]
\centering
\includegraphics[width=0.98\columnwidth]{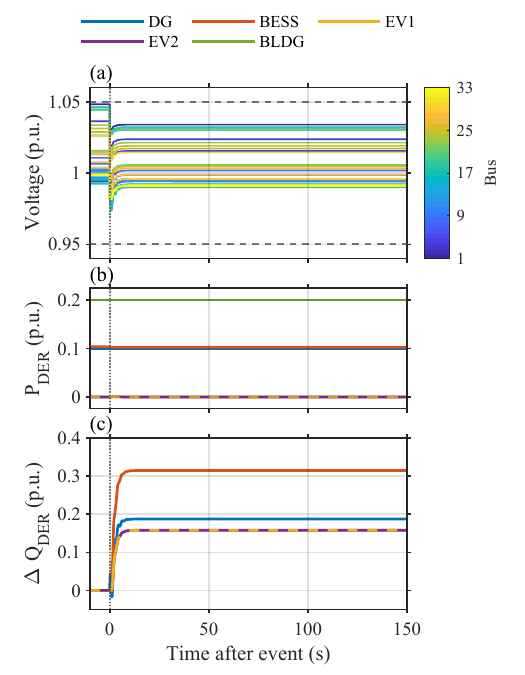}
\caption{Case~3 voltage support response: (a) bus voltages; (b) DER active powers; (c) DER reactive power deviations. The DG trace uses the filtered reactive power measurement; numerical metrics and constraint checks use the raw terminal outputs.}
\label{fig:voltage}
\end{figure}

The line trip event causes an immediate voltage drop throughout the feeder, with larger deviations at downstream buses in Fig.~\ref{fig:voltage}(a). The minimum voltage occurs at bus~18 and reaches 0.9737~p.u. at 0.869~s after the event, before the reactive power responses have fully developed. As the DERs increase their reactive power injections, the bus voltages recover from the initial drop and settle at new values within the prescribed limits. These values differ from those before the event because the increase in the equivalent grid resistance persists.

\FloatBarrier

The reactive power responses in Fig.~\ref{fig:voltage}(c) show how C-MPC allocates voltage support among the available DERs. The BESS provides the fastest and largest increase in reactive power, while the DG response develops more gradually and the two EV aggregations provide smaller contributions. The total DER reactive power support reaches a maximum of 0.8191 p.u. This allocation reflects the joint optimization of DER setpoints, accounting for their response dynamics, capability limits, and network operating constraints. 

This case also illustrates the separation between voltage support and the active power services. With PFC and AGC disabled, the active power outputs in Fig.~\ref{fig:voltage}(b) remain close to their values before the event, and the reactive power outputs adjust to the changed network condition. All optimization problems remain feasible during the 150~s following the event, with no constraint violations.

\section{Conclusions}\label{sec:conclusion}
In this paper, we have proposed a multi-rate MPC framework for coordinated PFC, AGC, and VC in active DNs. The formulation combines service setpoints at different update rates with physical DER responses and aggregate power--energy flexibility in a single optimization problem. Closed-loop simulations demonstrate that joint coordination can reduce control effort with frequency regulation comparable to the hierarchical benchmark. The results also show that multi-rate control is well suited to coordinating ancillary services across different time scales. In addition, the service update periods and shooting interval can be selected to balance frequency regulation, control effort, and computational requirements. Future work will consider uncertainty in DER availability and service requirements, communication delays, comparisons with additional control architectures, and validation on larger DNs.

\bibliographystyle{ieeetr}
\bibliography{reference}

@article{olivares2014,
  title = {Trends in microgrid control},
  author = {Olivares, D. E. and Mehrizi-Sani, A. and Etemadi, A. H. and Ca\~nizares, C. A. and Iravani, R. and Kazerani, M. and Hajimiragha, A. H. and Gomis-Bellmunt, O. and Saeedifard, M. and Palma-Behnke, R. and Jim\'enez-Est\'evez, G. A. and Hatziargyriou, N. D.},
  journal = {IEEE Transactions on Smart Grid},
  volume = {5},
  number = {4},
  pages = {1905--1919},
  year = {2014}
}

@article{karagiannopoulos2020,
  title = {Active distribution grids offering ancillary services in islanded and grid-connected mode},
  author = {Karagiannopoulos, S. and Gallmann, J. and Gonzalez Vay\'a, M. and Aristidou, P. and Hug, G.},
  journal = {IEEE Transactions on Smart Grid},
  volume = {11},
  number = {1},
  pages = {623--633},
  year = {2020}
}

@article{parisio2014,
  title = {A model predictive control approach to microgrid operation optimization},
  author = {Parisio, A. and Rikos, E. and Glielmo, L.},
  journal = {IEEE Transactions on Control Systems Technology},
  volume = {22},
  number = {5},
  pages = {1813--1827},
  year = {2014}
}

@article{kontis2021,
  title = {Provision of primary frequency response as ancillary service from active distribution networks to the transmission system},
  author = {Kontis, E. O. and Rodriguez del Nozal, A. and Mauricio, J. M. and Demoulias, C. S.},
  journal = {IEEE Transactions on Smart Grid},
  volume = {12},
  number = {6},
  pages = {4971--4982},
  year = {2021}
}

@inproceedings{tang2018,
  title = {Distributed control of active distribution networks for frequency support},
  author = {Tang, Z. and Liu, T. and Zhang, C. and Zheng, Y. and Hill, D. J.},
  booktitle = {Power Systems Computation Conference (PSCC)},
  pages = {1--7},
  year = {2018}
}

@article{rousis2021,
  title = {Provision of voltage ancillary services through enhanced {TSO--DSO} interaction and aggregated distributed energy resources},
  author = {Oulis Rousis, A. and Tzelepis, D. and Pipelzadeh, Y. and Strbac, G. and Booth, C. D. and Green, T. C.},
  journal = {IEEE Transactions on Sustainable Energy},
  volume = {12},
  number = {2},
  pages = {897--908},
  year = {2021}
}

@article{valverde2019,
  title = {Coordination of distributed reactive power sources for voltage support of transmission networks},
  author = {Valverde, G. and Shchetinin, D. and Hug-Glanzmann, G.},
  journal = {IEEE Transactions on Sustainable Energy},
  volume = {10},
  number = {3},
  pages = {1544--1553},
  year = {2019}
}

@article{karagiannopoulos2021,
  title = {Active distribution grids providing voltage support: The {Swiss} case},
  author = {Karagiannopoulos, S. and Mylonas, C. and Aristidou, P. and Hug, G.},
  journal = {IEEE Transactions on Smart Grid},
  volume = {12},
  number = {1},
  pages = {268--278},
  year = {2021}
}

@article{escobar2022,
  title = {Coordination of {DERs} and flexible loads to support transmission voltages in emergency conditions},
  author = {Escobar, F. and Viquez, J. M. and Garcia, J. and Aristidou, P. and Valverde, G.},
  journal = {IEEE Transactions on Sustainable Energy},
  volume = {13},
  number = {3},
  pages = {1344--1355},
  year = {2022}
}

@book{kundur1994,
  title = {Power System Stability and Control},
  author = {Kundur, P.},
  publisher = {New York: McGraw-Hill},
  volume = {7},
  year = {1994}
}

@article{zhao2014,
  title = {Design and stability of load-side primary frequency control in power systems},
  author = {Zhao, C. and Topcu, U. and Li, N. and Low, S. H.},
  journal = {IEEE Transactions on Automatic Control},
  volume = {59},
  number = {5},
  pages = {1177--1189},
  year = {2014}
}

@article{shen2025,
  title = {Hierarchical coordinated automatic generation control involving distributed energy resources: An aggregation approach},
  author = {Shen, Y. and Wu, W. and Wang, Q. and Li, Y. and Li, S. and Teng, W.},
  journal = {IEEE Transactions on Smart Grid},
  volume = {16},
  number = {2},
  pages = {1379--1391},
  year = {2025}
}

@article{zhang2017,
  title = {Evaluation of achievable vehicle-to-grid capacity using aggregate {PEV} model},
  author = {Zhang, H. and Hu, Z. and Xu, Z. and Song, Y.},
  journal = {IEEE Transactions on Power Systems},
  volume = {32},
  number = {1},
  pages = {784--794},
  year = {2017}
}

@article{hao2015,
  title = {Aggregate flexibility of thermostatically controlled loads},
  author = {Hao, H. and Sanandaji, B. M. and Poolla, K. and Vincent, T. L.},
  journal = {IEEE Transactions on Power Systems},
  volume = {30},
  number = {1},
  pages = {189--198},
  year = {2015}
}

@article{zhao2017,
  title = {A geometric approach to aggregate flexibility modeling of thermostatically controlled loads},
  author = {Zhao, L. and Zhang, W. and Hao, H. and Kalsi, K.},
  journal = {IEEE Transactions on Power Systems},
  volume = {32},
  number = {6},
  pages = {4721--4731},
  year = {2017}
}

@article{guerrero2011,
  title = {Hierarchical control of droop-controlled {AC} and {DC} microgrids---A general approach toward standardization},
  author = {Guerrero, J. M. and Vasquez, J. C. and Matas, J. and de Vicu\~na, L. G. and Castilla, M.},
  journal = {IEEE Transactions on Industrial Electronics},
  volume = {58},
  number = {1},
  pages = {158--172},
  year = {2011}
}

@article{scattolini2009,
  title = {Architectures for distributed and hierarchical model predictive control---A review},
  author = {Scattolini, R.},
  journal = {Journal of Process Control},
  volume = {19},
  number = {5},
  pages = {723--731},
  year = {2009}
}

@article{baran1989,
  title = {Network reconfiguration in distribution systems for loss reduction and load balancing},
  author = {Baran, M. E. and Wu, F. F.},
  journal = {IEEE Transactions on Power Delivery},
  volume = {4},
  number = {2},
  pages = {1401--1407},
  year = {1989}
}

@article{teng2003,
  title = {A direct approach for distribution system load flow solutions},
  author = {Teng, J.-H.},
  journal = {IEEE Transactions on Power Delivery},
  volume = {18},
  number = {3},
  pages = {882--887},
  year = {2003}
}

@article{chaspierre2020,
  title = {An active distribution network equivalent derived from large-disturbance simulations with uncertainty},
  author = {Chaspierre, G. and Denis, G. and Panciatici, P. and Van Cutsem, T.},
  journal = {IEEE Transactions on Smart Grid},
  volume = {11},
  number = {6},
  pages = {4749--4759},
  year = {2020}
}

@article{fulgencio2020,
  title = {Aggregated dynamic model of active distribution networks for large voltage disturbances},
  author = {Fulg\^encio, N. and Moreira, C. and Carvalho, L. and Pe\c{c}as Lopes, J.},
  journal = {Electric Power Systems Research},
  volume = {178},
  pages = {106006},
  year = {2020}
}

@article{namor2019,
  title = {Control of battery storage systems for the simultaneous provision of multiple services},
  author = {Namor, E. and Sossan, F. and Cherkaoui, R. and Paolone, M.},
  journal = {IEEE Transactions on Smart Grid},
  volume = {10},
  number = {3},
  pages = {2799--2808},
  year = {2019}
}

@inproceedings{li2023,
  title = {Unlocking energy flexibility from thermal inertia of buildings: A robust optimization approach},
  author = {Li, Y. and Yorke-Smith, N. and Keviczky, T.},
  booktitle = {IEEE Conference on Decision and Control (CDC)},
  pages = {2555--2562},
  year = {2023}
}

@inproceedings{lofberg2004,
  title = {{YALMIP}: A toolbox for modeling and optimization in {MATLAB}},
  author = {L\"ofberg, J.},
  booktitle = {IEEE International Symposium on Computer Aided Control Systems Design},
  year = {2004}
}

@book{milano2010,
  title = {Power System Modelling and Scripting},
  author = {Milano, F.},
  publisher = {Springer Science \& Business Media},
  year = {2010}
}

@article{pertl2019,
  title = {An equivalent time-variant storage model to harness {EV} flexibility: Forecast and aggregation},
  author = {Pertl, M. G. and Carducci, F. and Tabone, M. D. and Marinelli, M. and Kiliccote, S. and Kara, E. C.},
  journal = {IEEE Transactions on Industrial Informatics},
  volume = {15},
  number = {4},
  pages = {1899--1910},
  year = {2019}
}

@article{barot2017,
  title = {A concise, approximate representation of a collection of loads described by polytopes},
  author = {Barot, S. and Taylor, J. A.},
  journal = {International Journal of Electrical Power \& Energy Systems},
  volume = {84},
  pages = {55--63},
  year = {2017}
}

@article{wen2024,
  title = {Stochastic modeling for the aggregated flexibility of distributed energy resources},
  author = {Wen, Y. and Guo, Y. and Hu, Z. and Hug, G.},
  journal = {Electric Power Systems Research},
  volume = {234},
  pages = {110628},
  year = {2024}
}

@article{hanif2022,
  title = {Multi-service battery energy storage system optimization and control},
  author = {Hanif, S. and Alam, M. J. E. and Roshan, K. and Bhatti, B. A. and Bedoya, J. C.},
  journal = {Applied Energy},
  volume = {311},
  pages = {118614},
  year = {2022}
}

@article{carini1990,
  title = {Multirate self-tuning predictive control with application to binary distillation column},
  author = {Carini, P. and Micheli, R. and Scattolini, R.},
  journal = {International Journal of Systems Science},
  volume = {21},
  number = {1},
  pages = {51--64},
  year = {1990}
}

@article{heidarinejad2011,
  title = {Multirate {Lyapunov}-based distributed model predictive control of nonlinear uncertain systems},
  author = {Heidarinejad, M. and Liu, J. and Mu\~noz de la Pe\~na, D. and Davis, J. F. and Christofides, P. D.},
  journal = {Journal of Process Control},
  volume = {21},
  number = {9},
  pages = {1231--1242},
  year = {2011}
}

@inproceedings{gorbani2025,
  title = {Unified multi-rate model predictive control for a jet-powered humanoid robot},
  author = {Gorbani, D. and L'Erario, G. and Mohamed, H. A. O. and Pucci, D.},
  booktitle = {IEEE-RAS 24th International Conference on Humanoid Robots (Humanoids)},
  pages = {744--751},
  year = {2025}
}

@article{ravi2023,
  title = {Stochastic strategic participation of active distribution networks with high-penetration {DERs} in wholesale electricity markets},
  author = {Ravi, A. and Bai, L. and Cecchi, V. and Ding, F.},
  journal = {IEEE Transactions on Smart Grid},
  volume = {14},
  number = {2},
  pages = {1515--1527},
  year = {2023},
}

@article{miah2026,
  title = {Providing ancillary services by aggregated distributed energy resources of active distribution systems for the upstream grid: A critical review},
  author = {Miah, M. S. and Shah, R. and Aslam, M. U. and Amjady, N. and Walker, S. and Terzija, V. and Dehghan, S. and Mithulananthan, N.},
  journal = {Renewable and Sustainable Energy Reviews},
  volume = {233},
  pages = {116858},
  year = {2026},
}

@inproceedings{milano2018,
  title = {Foundations and challenges of low-inertia systems},
  author = {Milano, F. and D{\"o}rfler, F. and Hug, G. and Hill, D. J. and Verbi{\v{c}}, G.},
  booktitle = {2018 Power Systems Computation Conference (PSCC)},
  pages = {1--25},
  year = {2018},
}

@article{gutierrez2021,
  title = {Quantifying the effects of medium voltage--low voltage distribution network constraints and distributed energy resource reactive power capabilities on aggregators},
  author = {Gutierrez-Lagos, L. and Petrou, K. and Ochoa, L. F.},
  journal = {IET Generation, Transmission \& Distribution},
  volume = {15},
  number = {14},
  pages = {2019--2032},
  year = {2021},
}

@article{dozein2021,
  title = {Simultaneous provision of dynamic active and reactive power response from utility-scale battery energy storage systems in weak grids},
  author = {Dozein, M. G. and Gomis-Bellmunt, O. and Mancarella, P.},
  journal = {IEEE Transactions on Power Systems},
  volume = {36},
  number = {6},
  pages = {5548--5557},
  year = {2021},
}

@inproceedings{dhople2015,
  title = {Linear approximations to {AC} power flow in rectangular coordinates},
  author = {Dhople, S. V. and Guggilam, S. S. and Chen, Y. C.},
  booktitle = {2015 53rd Annual Allerton Conference on Communication, Control, and Computing (Allerton)},
  pages = {211--217},
  year = {2015},
}

@article{rao2003,
  title = {Constrained state estimation for nonlinear discrete-time systems: Stability and moving horizon approximations},
  author = {Rao, C. V. and Rawlings, J. B. and Mayne, D. Q.},
  journal = {IEEE Transactions on Automatic Control},
  volume = {48},
  number = {2},
  pages = {246--258},
  year = {2003},
}

@article{liu2021,
  title = {Dynamic state estimation for power system control and protection},
  author = {Liu, Y. and Singh, A. K. and Zhao, J. and Meliopoulos, A. P. and Pal, B. and Ariff, M. A. M. and Van Cutsem, T. and Glavic, M. and Huang, Z. and Kamwa, I. and Mili, L. and Mir, A. S. and Taha, A. and Terzija, V. and Yu, S.},
  journal = {IEEE Transactions on Power Systems},
  volume = {36},
  number = {6},
  pages = {5909--5921},
  year = {2021},
}

@article{cai2026,
  title = {Toward socio-techno-economic power systems with demand-side flexibility},
  author = {Cai, H. and Bellizio, F. and Guo, Y. and Humbert, G. and Montazeri, M. and Rousseau, J. and Brandes, M. and Chatterjee, A. and Gattiglio, A. and von Krannichfeldt, L. and Thrampoulidis, E. and Behrunani, V. N. and Strbac, G. and Heer, P.},
  journal = {Cell Reports Physical Science},
  volume = {7},
  pages = {103568},
  year = {2026},
}

@article{rebours2007,
  author = {Y. G. Rebours and D. S. Kirschen and M. Trotignon and S. Rossignol},
  title = {A Survey of Frequency and Voltage Control Ancillary Services---Part I: Technical Features},
  journal = {IEEE Transactions on Power Systems},
  volume = {22},
  number = {1},
  pages = {350--357},
  year = {2007},
}

@article{kerci2021,
  title = {On the impact of discrete secondary controllers on power system dynamics},
  author = {K{\"e}r{\c{c}}i, T. and Murad, M. A. A. and Dassios, I. and Milano, F.},
  journal = {IEEE Transactions on Power Systems},
  volume = {36},
  number = {5},
  pages = {4400--4409},
  year = {2021},
}

\appendix
\label{sec:benchmark}
This appendix describes the H-MPC architecture used for comparison in Section V. H-MPC uses an upper layer MPC for AGC scheduling and a fast layer with P--f and Q--V droop for PFC and VC \cite{scattolini2009,guerrero2011}. At each scheduled AGC update $m\bmod q=0$, we optimize the AGC setpoint increments on the common time grid using the same distribution system and demand-side flexibility models as C-MPC. The decision vector is
\begin{equation}
\bm\Omega_t^a:=\bigl(
\Delta\mathbf u^{\AGC}_{k|t},\eta^{\AGC}_{k|t}
\bigr)_{k\in\mathcal A_m}.
\label{eq:upper_variables}
\end{equation}

Let $\mathcal X_t^{\textrm{PE},a}$ collect the response models, DER capability limits, and aggregate power--energy flexibility bounds used in the upper layer, all evaluated at $k\in\mathcal K$ with the current initial states. The DER responses are predicted with sampling period $T_s$, while the AGC setpoints can change only at the steps in $\mathcal A_m$. The PFC and VC setpoints are held at their currently applied values in the upper layer prediction, although their physical responses may continue to evolve. This set also includes the AGC setpoint recursions in \eqref{eq:general_increment} and the update constraints in \eqref{eq:general_update_constraint}, with $\mathbf u^{\AGC}_{0|t}=\mathbf u^{\AGC}(t^-)+\Delta\mathbf u^{\AGC}_{0|t}$. The remaining predicted quantities are determined by these setpoints and the current initial states. The superscript $a$ identifies quantities in the upper layer prediction. The upper layer problem is
\begin{subequations}\label{eq:upper_mpc}
\begin{align}
\bm{\mathcal P}_t^a:\quad \;\underset{\bm\Omega_t^a}{\min}\quad
&\sum_{k\in\mathcal A_m}
\|\Delta\mathbf u^{\AGC}_{k|t}\|_{\mathbf W_{\AGC}}^2
+\gamma\sum_{k\in\mathcal A_m}|\eta^{\AGC}_{k|t}|,
&\label{eq:upper_problem}\\
\text{subject to}\quad
&(\mathbf u^{\AGC},\mathbf x^a,\mathbf p^a,
  \mathbf q^a,\mathbf E^a)
\in\mathcal X_t^{\textrm{PE},a},
\label{eq:upper_device_constraints}\\
&\begin{bmatrix}
\Delta\mathbf v^a_{k|t}\\
\Delta\mathbf i^a_{k|t}
\end{bmatrix}
=\mathbf F_t
\begin{bmatrix}
\Delta\mathbf p^{\net,a}_{k|t}\\
\Delta\mathbf q^{\net,a}_{k|t}
\end{bmatrix},
\quad k\in\mathcal K,
\label{eq:upper_network_model}\\
&\mathbf g_n\bigl(
\hat{\mathbf V}_t+\Delta\mathbf v^a_{k|t},
\hat{\mathbf I}_t+\Delta\mathbf i^a_{k|t}
\bigr)\leq\mathbf0,
\quad k\in\mathcal K,\notag\\
&\label{eq:upper_network_constraints}\\
&y^{\AGC}_{k|t}
=\rho_t^{\AGC}+\eta^{\AGC}_{k|t},
\quad k\in\mathcal A_m.
\label{eq:upper_agc_matching}
\end{align}
\end{subequations}
Here, $\Delta\mathbf p^{\net,a}_{k|t}$ and $\Delta\mathbf q^{\net,a}_{k|t}$ denote the total net injection changes predicted by the upper layer. They include the AGC response and the physical responses associated with the held PFC and VC setpoints. The signed mismatch $\eta^{\AGC}_{k|t}\in\R$ is evaluated at the end of the first prediction step following the scheduled AGC update, consistent with the centralized formulation. We apply the first optimized AGC setpoint and hold it until the next scheduled AGC update. The remaining setpoint sequence is used to predict the DER response and evaluate constraints over the horizon. At the next scheduled update, we update the operating point and solve problem~\eqref{eq:upper_mpc} again.

With the AGC allocation fixed, the fast layer computes the unconstrained PFC and VC setpoints as
\begin{subequations}\label{eq:hierarchical_targets}
\begin{align}
u_{d,t}^{\PFC,H} &= K_{P,d}^H e_t^f,\qquad d\in\mathcal D, \label{eq:hierarchical_pfc_target}\\
u_{d,t}^{\VC,H} &= K_{V,d}^H\,e_t^{\VC},\qquad d\in\mathcal D_Q, \label{eq:hierarchical_vc_target}\\
K_{P,d}^H &= K_P\,\alpha_{P,d}^H,\qquad K_{V,d}^H=K_V\,\alpha_{Q,d}^H, \label{eq:hierarchical_gains}\\
\alpha_{P,d}^H
&=s_{P,d}\frac{P_d^{\mathrm{cap}}}{\sum_{j\in\mathcal D}P_j^{\mathrm{cap}}},
\qquad d\in\mathcal D, \label{eq:hierarchical_p_participation}\\
\alpha_{Q,d}^H
&=\frac{Q_d^{\mathrm{cap}}}{\sum_{j\in\mathcal D_Q}Q_j^{\mathrm{cap}}},
\qquad d\in\mathcal D_Q, \label{eq:hierarchical_q_participation}
\end{align}
\end{subequations}
where $P_d^{\mathrm{cap}}>0$ and $Q_d^{\mathrm{cap}}>0$ denote the capacities used for active and reactive power sharing at the current update. For the EV aggregations, these capacities vary with vehicle availability. The vectors $\mathbf u_t^{\PFC,H}$ and $\mathbf u_t^{\VC,H}$ collect the corresponding PFC and VC setpoints. The participation factors are proportional to these capacities, following the droop design based on ratings in \cite{guerrero2011}. The factor $s_{P,d}$ accounts for the active power sign convention for upstream support, giving $\sum_{d\in\mathcal D}s_{P,d}\alpha_{P,d}^H=1$ and $\sum_{d\in\mathcal D_Q}\alpha_{Q,d}^H=1$.

The droop setpoints are adjusted to satisfy local device limits. The DG reactive power limit is determined from its active power, and the setpoints of the EV and TCL aggregations are limited using their power and energy bounds for the next fast control interval. The upper layer AGC allocation remains fixed during these adjustments. This local calculation does not use feeder sensitivities or impose feeder voltage and branch current constraints.



\end{document}